\documentclass[12pt]{article}
\pdfoutput=1
\usepackage{subfigure}
\usepackage{amssymb,amsmath}
\usepackage{graphicx}
\usepackage{color}
\usepackage[colorlinks=true
,urlcolor=blue
,citecolor=blue
,linkcolor=blue
,pagecolor=blue
,linktocpage=true
,pdfproducer=medialab
]{hyperref}
\usepackage[a4paper]{geometry}

\usepackage{placeins}

\makeatletter \renewcommand{\@dotsep}{10000} \makeatother

\mathchardef\mhyphen="2D

\newcommand{\beq}{\begin{equation}}
\newcommand{\eeq}{\end{equation}}
\newcommand{\bea}{\begin{eqnarray}}
\newcommand{\eea}{\end{eqnarray}}

\usepackage[nodisplayskipstretch]{setspace}

\begin{document}

\begin{titlepage}
\pagestyle{empty}

\vspace*{0.2in}
\begin{center}
{\Large \bf    Just a Scalar in THDM
  }\\
\vspace{1cm}
{\bf  B\"{u}\c{s}ra Ni\c{s}$^{a,}$\footnote{E-mail: 501507008@ogr.uludag.edu.tr},  Ali \c{C}i\c{c}i$^{a,}$\footnote{E-mail: 501507007@ogr.uludag.edu.tr}, Zerrin K\i rca $^{a,}$\footnote{E-mail: zkirca@uludag.edu.tr} and
Cem Salih $\ddot{\rm U}$n$^{a,}$\footnote{E-mail: cemsalihun@uludag.edu.tr}}
\vspace{0.5cm}

{\it $^a$Department of Physics, Uluda\~{g} University, TR16059 Bursa, Turkey
}

\end{center}

\vspace{0.5cm}
\begin{abstract}
\noindent
We review the THDM model in which one of the Higgs doublets does not develop a vacuum expectation value. In this case, the Higgs fields with zero VEV does not contribute to the physical masses of the SM particles, and hence, its interactions with the SM particles can have more freedom than in the case of usual considerations on the THDM models. We show that the stability of the Higgs potential minima can be maintained. All the Higgs boson masses are found lighter than about 300 GeV in the low scale spectrum when $v_{1}=0$. Such light mass scales are in the detectable regime, and hence, the SM predictions and the experimental results are essential to be applied in the analyses. Especially the decay channels; $h\rightarrow W^{+}W^{-}$ and $h\rightarrow b\bar{b}$ exclude most of the solutions, while it is still possible to realize a small region, which coincides with the SM predictions. We highlight that it is also possible to realize an excess in $h\rightarrow \gamma\gamma$ decay channel, even one applies the constraints from the Higgs boson decays into $W^{+}W^{-}$ and $b\bar{b}$. In addition, if one assumes $H$ is the SM-like Higgs boson of mass about 125 GeV, the solutions with $m_{h}\lesssim 125$ GeV can be acceptable. In this case, the solutions can still be realized consistent with the SM predictions; however, an excess in $H\rightarrow \gamma\gamma$ cannot be observed, while the implications for this channel in THDM can stay in the SM prediction rates at most. Even though the light masses of the Higgs bosons can be favored in resolution to discrepancy between the Standard Model and the experiment in the muon anomalous magnetic moment measurements, we find that it is not possible to accommodate such a resolution with $BR(b\rightarrow s\gamma)$ results simultaneously.

\end{abstract}

\end{titlepage}



\section{Introduction}
\label{sec:Intro}

The discovery of the Higgs boson by the ATLAS \cite{Aad:2012tfa} and the CMS \cite{Chatrchyan:2012xdj} collaborations has undoubtedly started a new era in the high energy particle physics. The experimental results and analyzes have revealed that the observations on the Higgs boson overlap with the Standard Model (SM) predictions in a very good agreement, despite observed excesses in some decay channels of the Higgs boson such as those with final states of two photons \cite{CMS:ril} and two pairs of charged leptons \cite{Chatrchyan:2013mxa}. Although its predictions have been mostly confirmed, the SM is problematic in the Higgs boson, since it suffers from the infinitive loop corrections to the Higgs boson mass (gauge hierarchy problem) \cite{Gildener:1976ai} and also the absolute stability of the Higgs potential \cite{hinstability}. 

Such problems arise within the SM as a result of the fact that the SM gauge group cannot protect the Higgs boson mass, while the fermion masses are safe from such infinitive loop corrections.  Besides, such a gauge group also does not constrain the number of the Higgs bosons, and hence one can extend the SM by adding new scalar fields to the theory. In this sense, even though there is a sort of arbitrariness in the number of scalar doublet fields, it can be restricted by imposing the gauge coupling unification at some high energy scale. A theory with two Higgs doublets can unify the gauge couplings of the SM, they deviate from the unification when the number of the Higgs doublets is more than two \cite{Baer:2006rs}. 

Even though the two Higgs doublet models (THDM) \cite{Branco:2011iw} does not provide a resolution to the gauge hierarchy problem in its minimal setup, it can be considered as a low scale projection of some larger models such as supersymmetry, THDM extended with vector-like fermions \cite{Aguilar-Saavedra:2013qpa} etc. In this context, we will consider THDM in our paper in its minimal version, in which the SM is extended only by adding another Higgs doublet with the same quantum numbers as the SM Higgs boson. Even such a simple extension significantly raises the number of physical Higgs boson states, since it yields two CP-even, one CP-odd and two charged Higgs bosons in its particle spectrum. Assuming one of the CP-even Higgs bosons is responsible for the experimental observations about a scalar boson of mass about 125 GeV, the other Higgs bosons can significantly alter the Higgs phenomenology (see for instance \cite{Un:2016hji}). For example, the excesses in $h\rightarrow \gamma \gamma$ have also been observed at $m_{\gamma\gamma} \sim 137$ and $145$ GeV \cite{CMS:ril}, in addition to that observed at $m_{\gamma\gamma} \sim 125$ GeV. While the excess at $m_{\gamma\gamma}\sim 125$ GeV can be explained with the SM-like Higgs boson, the other excesses require extra Higgs bosons, which are not included in the SM. We also investigate if the extra Higgs bosons of THDM can fit into the experimental observations in light of such excesses.

Accordingly to the abundance of the Higgs bosons, the Higgs potential in THDM is also more complicated than the SM, since it includes the mixing between the two Higgs doublets as well as the self-interactions. Such a potential yields two vacuum expectation values (VEVs), one for each Higgs doublet, and the electroweak symmetry breaking (EWSB) significantly constrains these VEVs in order to yield a consistent physical implications. For instance, all the Higgs boson mass states and their couplings to the SM particles are stringently restricted by the fermion and gauge boson masses. It is interesting here to note that if one imposes a $Z_{2}$ symmetry, it is possible to have a stable minimum for the Higgs potential, when one of the Higgs doublet does not develop a non-zero VEV. In this case,  the Higgs doublet with zero VEV can be kept from constraints by the fermion and gauge boson masses, while it interferes in the interactions with the SM particles, as it couples to them at tree-level. 

In addition to the Higgs potential, the Yukawa interactions between the Higgs fields and SM fermions can be given in a general form as follows;

\begin{equation*}\hspace{-3.0cm}
-\mathcal{L} _{yuk} = \xi_{ij}^{U} \bar{Q}_{iL} \tilde{\Phi}_{1}U_{jR}
+ \xi_{ij}^{D} \bar{Q}_{iL} \Phi_{1} D_{jR}+  \xi_{ij}^{D} \bar{L}_{iL} \Phi_{1} E_{jR}
\end{equation*}
\begin{equation}
+ \eta_{ij}^{U} \bar{Q}_{iL} \tilde{\Phi}_{2} U_{jR}
+ \eta_{ij}^{D} \bar{Q}_{iL} \Phi_{2} D_{jR}
+ \eta_{ij}^{E} \bar{l}_{iL} \Phi_{2} E_{jR}
+h.c.
\end{equation}
where we adopted the usual notation for the fields as $Q$ and $L$ denote the left-handed quark and lepton fields respectively, while $U$, $D$ and $E$ stand for the right-handed up-type, down-type quarks and leptons with family indices $i,j=1,2,3$, and $\xi$ and $\eta$ stand for the Yukawa couplings between the SM fermions and the Higgs doublets $\Phi_{1}$ and $\Phi_{2}$ respectively. Similarly, a $Z_{2}$ symmetry can be applied to the Yukawa Lagrangian. Indeed, different types of THDM are classified with the applied $Z_{2}$ symmetry to its Yukawa sector, and Table \ref{table1} summarizes the different types of THDM and their $Z_{2}$ symmetries.

\begin{table}[h]
\centering
\setstretch{1.5}
\scalebox{0.8}{
\tabcolsep7pt\begin{tabular}{|c|c|c|}
\hline
  {Type I}  & Type II  & Type III    \\
\hline
$\Phi_{1} \rightarrow \Phi_{1}$,  $\Phi_{2} \rightarrow -\Phi_{2}$ & $\Phi_{1} \rightarrow \Phi_{1}$, $\Phi_{2} \rightarrow -\Phi_{2}$ &
\\
$D_{jR} \rightarrow -D_{jR}$,  $ \ U_{jR} \rightarrow -U_{jR}$ &
$D_{jR} \rightarrow D_{jR}$,  $ \ U_{jR} \rightarrow -U_{jR}$ &
\\&&
\\
$\xi_{ij}^{U,D,E} = 0$ & $\xi_{ij}^{U} = 0$,  $\eta_{ij}^{D,E} = 0$  &
\\&&
\\
$M_{U,D,E} = \dfrac{v_{2} \eta^{U,D,E}}{\sqrt{2}}$ & $M_{D,E} = \dfrac{v_{1} \xi^{D,E}}{\sqrt{2}}$ and $M_U = \dfrac{v_{2} \eta^U}{\sqrt{2}}$ & $M_{U,D,E} = \dfrac{v_{1} \eta^{U,D,E} + v_{2} \xi^{U,D,E} }{\sqrt{2}}$  \\ && \\
\hline
\end{tabular}
}
\caption{The $Z_{2}$ symmetries and different types of THDM.}
\label{table1}
\end{table}

If the Yukawa Lagrangian is required to be invariant under a $Z_{2}$ symmetry as $D\rightarrow -D$ and $U\rightarrow -U$, it forbids the interactions between $\Phi_{1}$ and the SM fermions, and hence all the fermions acquire their masses only from $\Phi_{2}$. Such models of THDM are classified as Type-I. Similarly in a Yukawa Lagrangian symmetric under $D\rightarrow D$ and $U\rightarrow -U$ transformations leads to interactions that $D$ and $E$ interact only with $\Phi_{1}$, while $U$ interacts only with $\Phi_{2}$. Such models are called Type-II. It should be noted that different types of quarks gain their masses from different Higgs doublets, and their masses are proportional to the VEVs of these Higgs fields. If one of the Higgs fields does not develop a non-zero VEV (say $v_{1}=0$), then Type-II models fail to satisfy the constraints from the fermion masses. Thus, even though the $Z_{2}$ symmetry can still be applied to the Higgs sector, the minima of the Higgs potential with $v_{1}=0$ cannot be considered physical. Finally, if there is no symmetry applied in the Yukawa Lagrangian, then each SM fermion can interact with both Higgs doublets, and this THDM framework is classified as Type-III.

As discussed in the beginning that we would focus on the case in which one of the Higgs doublets does not develop a non-zero VEV ($v_{1}=0$), and analyze its implications for the Higgs bosons and its interactions. In this context, since the solutions with $v_{1}=0$ cannot be consistent with the physical fermion masses, we rather consider THDM Type-III. The rest of the paper is the following: In Section \ref{sec:scan} we will summarize our scanning procedure in generating numerical data for THDM, and also we highlight the relevant experimental constraints. Section \ref{sec:Potential} discusses the Higgs potential and stability when one of the Higgs doublets does not develop a non-zero VEV. We also present our results for the mass spectrum of the Higgs bosons in this section. Then, we discuss our results in comparison to the SM predictions in Section \ref{sec:comparison}. Finally, we conclude and summarize our results in Section \ref{sec:conc}.

\section{Scanning procedure and Experimental Constraints}
\label{sec:scan}

We have employed SPheno 3.3.8 package \cite{Porod:2003um} obtained with SARAH 4.5.8 \cite{Staub:2008uz}. We modified the packages in order to obtain the numerical codes compatible THDM, in which one of the Higgs doublets does not develop a VEV. In the SPheno package, the gauge and Yukawa couplings along with the fermion and gauge boson masses are supplied as input, and the couplings of the Higgs doublets with the SM particles are calculated in terms of them. Since we have only one VEV, the couplings for the other Higgs doublets are needed to be provided in the input as well. The model is considered rather an effective model at the low scale, and we set the renormalization scale to 1 TeV. Above this scale, the physics might be described by a larger model. 

We generate the numerical data by scanning over the following set of the free parameters: $\eta^{U,D,E}, \mu_{1},\lambda_{i}$, where $\eta^{U,D,E}$ have already defined in the previous section. $\mu_{1}$ is the mass parameter for $\Phi_{1}$ and $\lambda_{i}$ are the couplings of the interactions between the Higgs fields including the self-interactions. $\mu_{1}$ and $\lambda_{i}$ will be discussed in details in Section \ref{sec:Potential}. In scanning we use our interface which employs Metropolis-Hasting algorithm described in \cite{Belanger:2009ti}. We also set the top quark mass to its central value as $m_{t}=173.3$ GeV \cite{Group:2009ad}. Note that the Higgs boson masses can differ by 1-2 GeV by variation within the $1-2\sigma$ uncertainty in top quark mass \cite{Gogoladze:2011aa}.

All points collected satisfy the EWSB constraints. After generating the data, we impose the experimental constraints to distract the solutions which are allowed by the experimental results. The first constraint comes from the Higgs boson such that the spectrum is required to yield at least one  scalar boson of mass about 125 GeV. We do not impose constraints from couplings and decays for such a scalar directly, but we analyze its decays separately. 

In addition to the Higgs boson, one of the most important constraint comes from the rare decays of $B-$meson, especially those occurring through the $b\rightarrow s\gamma$ decay. Even though such decays can happen only at loop levels, the experimental results for this decay (${BR}(b\rightarrow s\gamma) = (3.43\pm 0.22)\times 10^{-4}$ \cite{Amhis:2012bh}) coincides with the SM prediction (${BR}(b\rightarrow s\gamma) = (3.15\pm 0.23)\times 10^{-4}$ \cite{Misiak:2006zs}). Since the extra Higgs bosons in the spectrum, especially the charged ones, contribute to this decay process, they can result in a significant deviation from the SM values, and hence the solutions can easily be excluded by the experiment. The amplitude for $b\rightarrow s\gamma$ mediated with the charged Higgs boson can be given \cite{Deshpande:1987nr}

\begin{equation}
M_{H^{\pm}}^{b\rightarrow s\gamma}=\dfrac{G_{F}}{\sqrt{2}}\left(\dfrac{e}{16\pi^{2}} \right)\bar{s}\sigma_{\lambda\nu}q^{\nu}(1+\gamma_{5})b~m_{b}V_{tb}V_{ts}^{*}[G(\delta)+Q_{t}F(\delta)]\epsilon^{\lambda}
\end{equation}
where $G_{F}$ is the Fermi constant, $e$ is the electric charge, while $b,s$ denote spinors for the bottom and strange quarks, and $\epsilon^{\lambda}$ is the polarization of the photon. $m_{b}$ is the mass of the bottom quark, and $V_{tb},V_{ts}$ are the CKM matrix elements for the quarks shown in the subindices. $G(\delta)$ and $F(\delta)$ are the form factors induced at the loop \cite{Idarraga:2008zz}.

\begin{figure}[ht!]
\subfigure{\includegraphics[scale=1]{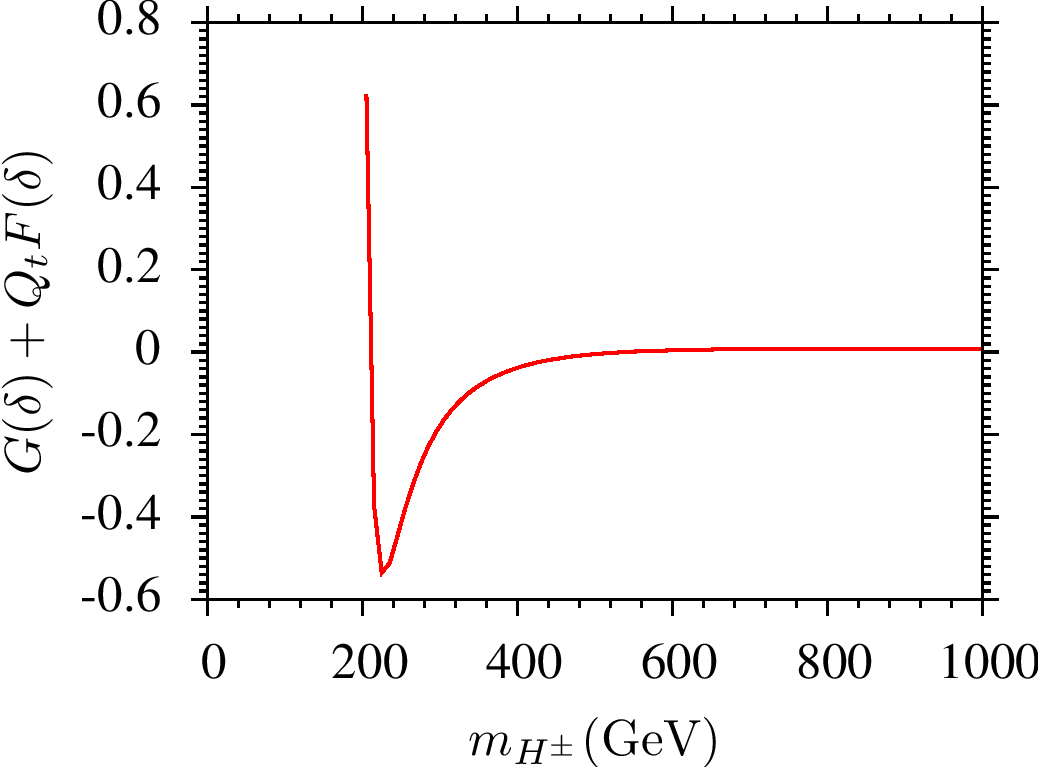}}
\subfigure{\includegraphics[scale=1]{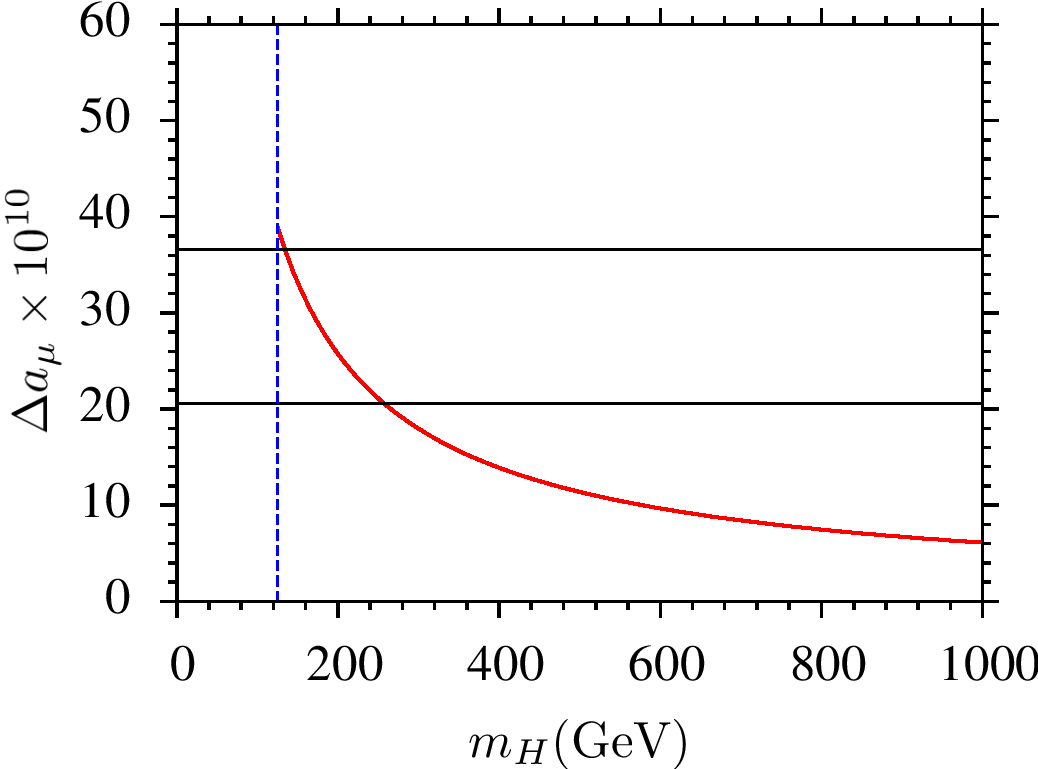}}
\caption{The contributions to $b\rightarrow s \gamma$ and muon $g-2$ from the extra Higgs bosons of THDM. We assume the couplings of these Higgs bosons to the SM particles are the same as those of the SM Higgs boson.}
\label{fig:constraints}
\end{figure} 

The agreement with the SM and the experimental results over ${\rm BR}(b\rightarrow s \gamma)$ provides a stringent constraint on the solutions that either the extra Higgs bosons negligibly couple to the SM particles, or they are so heavy that their masses significantly suppress the loop contributions. Another consideration over the solutions may come from the muon anomalous magnetic moment (hereafter muon $g-2$), which also induced at loop levels. The SM predictions for the muon $g-2$ represents the largest discrepancy between the SM and the experimental results. In this context, the resolution to the muon $g-2$ discrepancy rather requires light new particles, the extra Higgs bosons in THDM, which can significantly contribute to muon $g-2$. As a result, the solutions for resolution to the muon $g-2$ problem also bring a tension with the constraints from the $b\rightarrow s\gamma$ processes. If we assume that the extra Higgs bosons have the same couplings to the particles as those the SM Higgs boson has, then the contributions to the $b\rightarrow s\gamma$ process and muon $g-2$ restrict the masses of these Higgs bosons. Figure \ref{fig:constraints} represents the comparison between the contributions to $b\rightarrow s \gamma$ and muon $g-2$ from the extra Higgs bosons of THDM. We assume the couplings of these Higgs bosons to the SM particles are the same as those of the SM Higgs boson. As seen from the left panel, the Higgs boson can alter the implications for $b\rightarrow s\gamma$ significantly when $m_{H^{\pm}} \lesssim 400$ GeV. Such mass scales can be excluded, unless their couplings to the SM particles are negligible. On the other hand, as is seen from the right panel, the same mass scales are favored by the resolution to the muon $g-2$ discrepancy, and the contributions to muon $g-2$ decreases sharply when $m_{H} \gtrsim 400$ GeV.

Based on the discussion above, we apply the constraint from $b\rightarrow s\gamma$, since its experimental measurements are strict and sensitive. On the other hand, we do not require the solutions to resolve the muon $g-2$ discrepancy, and we accept only the solutions which do not worsen the SM predictions for the muon $g-2$. After all the constraints employed in our analyses can be summarized as follows:

\begin{equation}
\setstretch{1.5}
\begin{array}{rcl}
123  \leq & m_{h}~{\rm or}~m_{H} & \leq 127~{\rm GeV}\\
2.99\times 10^{-4} \leq & {\rm BR}(b\rightarrow s\gamma) & \leq 3.87\times 10^{-4}~(2\sigma) \\
0\leq & \Delta a_{\mu} &
\end{array}
\end{equation}
where $\Delta a_{\mu}\equiv a_{\mu}^{{\rm THDM}}-a_{\mu}^{{\rm SM}}$ represents the difference in muon $g-2$ predictions between THDM and the SM.

\section{The Higgs Potential and Masses}
\label{sec:Potential}

As mentioned in the beginning, the gauge symmetry of the SM does not prevent to add more Higgs fields, which can manage the electroweak symmetry sector together. In this section we will consider the Higgs potential in a general form, which is spanned with two Higgs doublets with the same quantum numbers  

\begin{equation}
\setstretch{1.5}
\Phi_{1}=\left( \begin{array}{c}
\phi_{1}^{+} \\ \phi_{1}^{0}
\end{array}   \right)~,\hspace{0.3cm} \Phi_{2}=\left( \begin{array}{c}
\phi_{2}^{+} \\ \phi_{2}^{0}
\end{array}   \right)
\label{fields}
\end{equation}

Neutral components of these fields can develop VEVs denoted as $v_{1}$ and $v_{2}$ respectively, which can be determined by minimizing the Higgs potential. The most general form of the Higgs potential can be written as

\begin{equation*}\hspace{-0.5cm}
V(\Phi_{1},\Phi_{2})= -\mu _{1}^{2} (\Phi_{1}^\dagger \Phi_{1}) 
-\mu _{2}^{2} (\Phi_{2}^\dagger \Phi_{2}) 
-\mu _{3}^{2} [\frac{1}{2} (\Phi_{1}^\dagger \Phi_{2}+\Phi_{2}^\dagger \Phi_{1})]
-\mu _{4}^{2} [-\frac{i}{2} (\Phi_{1}^\dagger \Phi_{2} - \Phi_{2}^\dagger \Phi_{1})]
\end{equation*}
\begin{equation*}\hspace{-2.1cm}
+\lambda _{1} (\Phi_{1}^\dagger \Phi_{1})^{2}
+\lambda _{2} (\Phi_{2}^\dagger \Phi_{2})^{2} 
+\lambda _{3} [\frac{1}{2} (\Phi_{1}^\dagger \Phi_{2}+\Phi_{2}^\dagger \Phi_{1})]^{2}
\end{equation*}
\begin{equation*}\hspace{2.3cm}
+\lambda _{4} [-\frac{i}{2} (\Phi_{1}^\dagger \Phi_{2} - \Phi_{2}^\dagger \Phi_{1})]^{2}
 +\lambda _{5} (\Phi_{1}^\dagger \Phi_{1}) (\Phi_{2}^\dagger \Phi_{2}) 
+\lambda _{6} (\Phi_{1}^\dagger \Phi_{1}) [\frac{1}{2} (\Phi_{1}^\dagger \Phi_{2}+\Phi_{2}^\dagger \Phi_{1})]
\end{equation*}
\begin{equation*}\hspace{-0.1cm}
 +\lambda _{7} (\Phi_{2}^\dagger \Phi_{2}) [\frac{1}{2} (\Phi_{1}^\dagger \Phi_{2}+\Phi_{2}^\dagger \Phi_{1})]
+\lambda _{8} (\Phi_{1}^\dagger \Phi_{1})  [-\frac{i}{2} (\Phi_{1}^\dagger \Phi_{2} - \Phi_{2}^\dagger \Phi_{1})]
\end{equation*}
\begin{equation}\hspace{2.3cm}
+\lambda _{9} (\Phi_{2}^\dagger \Phi_{2}) [-\frac{i}{2} (\Phi_{1}^\dagger \Phi_{2} - \Phi_{2}^\dagger \Phi_{1})]
+\lambda _{10} [\frac{1}{2} (\Phi_{1}^\dagger \Phi_{2}+\Phi_{2}^\dagger \Phi_{1})] [-\frac{i}{2} (\Phi_{1}^\dagger \Phi_{2} - \Phi_{2}^\dagger \Phi_{1})]~.
\label{Higgspot}
\end{equation}

The terms in Eq.(\ref{Higgspot}), except those with $\mu_{1}$, $\mu_{2}$, $\lambda_{1}$ and $\lambda_{2}$, mix the two Higgs fields each other, and hence the physical Higgs states involving two CP-even, one CP-odd and two charged Higgs boson mass eigenstates, emerge as linear superpositions of these fields. 

Even though THDM is the simplest extension of the SM, the number of free parameters are significantly raised, since all the parameters in Eq.(\ref{Higgspot}) are, in principle, free parameters. If one requires this potential to preserve some symmetries, then some of the terms vanish. For instance, if $\mu_{4}=\lambda_{8}=\lambda_{9}=\lambda_{10}=0$, then the potential remains invariant under the charge conjugation transformations (C-invariance). In addition, if we require the invariance under a $Z_{2}$ symmetry with $\Phi_{1}\rightarrow \Phi_{1}$ and $\Phi_{2}\rightarrow -\Phi_{2}$, then it leads to $\mu_{3}=\mu_{4}=\lambda_{6}=\lambda_{7}=\lambda_{8}=\lambda_{9}=\lambda_{10}=0$. Thus, the potential given in Eq.(\ref{Higgspot}) reduces to 

\begin{equation*}\hspace{-0.5cm}
V(\Phi_{1},\Phi_{2})= -\mu _{1}^{2} (\Phi_{1}^\dagger \Phi_{1})-\mu _{2}^{2} (\Phi_{2}^\dagger \Phi_{2})+\lambda _{1} (\Phi_{1}^\dagger \Phi_{1})^{2}
+\lambda _{2} (\Phi_{2}^\dagger \Phi_{2})^{2} 
+\lambda _{3} [\frac{1}{2} (\Phi_{1}^\dagger \Phi_{2}+\Phi_{2}^\dagger \Phi_{1})]^{2}
\end{equation*}
\begin{equation}\hspace{-3.3cm}
+\lambda _{4} [-\frac{i}{2} (\Phi_{1}^\dagger \Phi_{2} - \Phi_{2}^\dagger \Phi_{1})]^{2}
 +\lambda _{5} (\Phi_{1}^\dagger \Phi_{1}) (\Phi_{2}^\dagger \Phi_{2}) 
 \label{HiggspotZ2}
\end{equation}

Furthermore, the two of the remaining free parameters can be determined by the electroweak symmetry breaking. Although there are a number of different sets of solutions to the VEVs, one set is interesting in which $v_{1}=0$ and $v_{2}=v_{{\rm SM}}$, where $v_{{\rm SM}}$ is the VEV for the Higgs field predicted by the SM. Note that in each set of solutions, the condition $v_{1}^{2}+v_{2}^{2}=v_{{\rm SM}}^{2}$ has to be satisfied. 

In our study, we focus on the THDM framework, in which $\Phi_{1}$ does not develop a non zero VEV, and hence the model basically reduces to the SM extended with a scalar field. In this case, the scalar field with zero VEV can avoid from the constraints from the masses of the SM gauge bosons and fermions. The minimization of the Higgs potential yields the following tadpole equations

\begin{equation}
\setstretch{2.5}
\begin{array}{l}
-\mu_{1}^{2}v_{1}+\lambda_{1}v_{1}^{3}+\dfrac{1}{2}(\lambda_{3}+\lambda_{5})v_{1}v_{2}^{2}=0 \\
-\mu_{2}^{2}v_{2}+\lambda_{2}v_{2}^{3}+\dfrac{1}{2}(\lambda_{3}+\lambda_{5})v_{1}^{2}v_{2}=0~.
\end{array}
\end{equation}

\begin{figure}[ht!]\hspace{-2.0cm}
\subfigure{\includegraphics[scale=0.65]{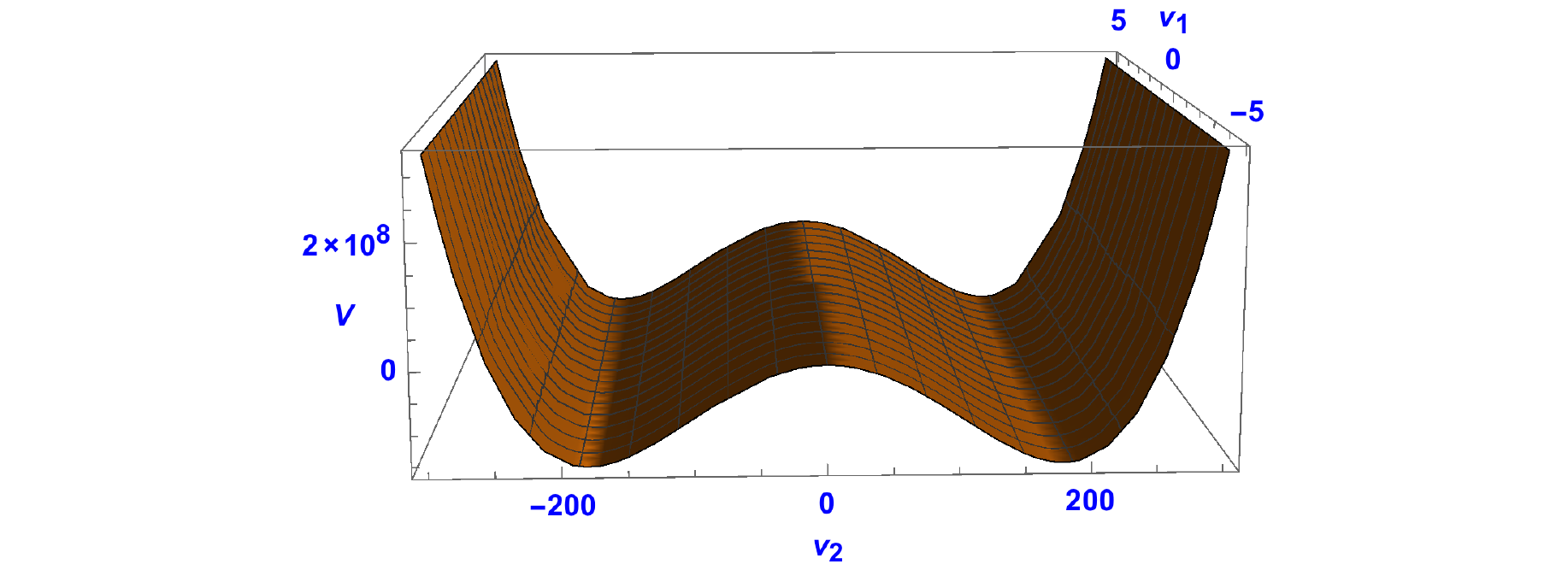}}%
\subfigure{\includegraphics[scale=0.5]{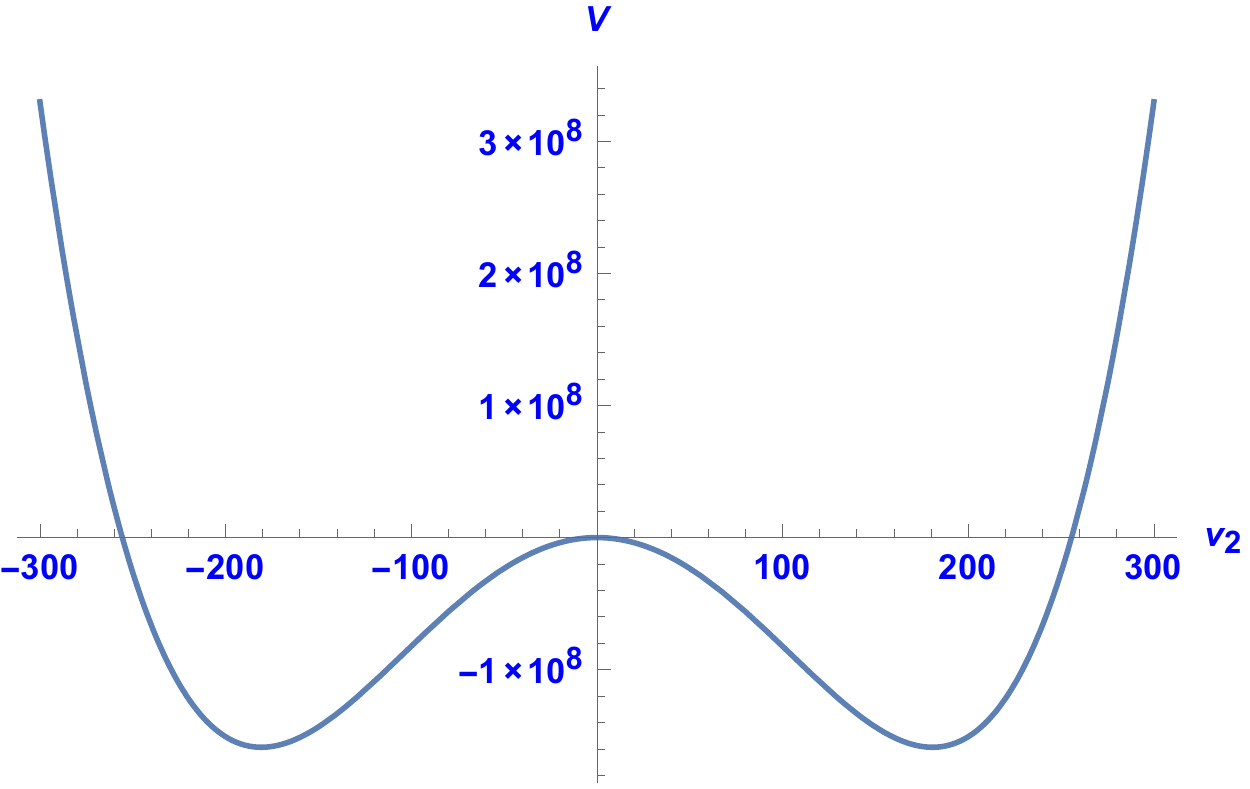}}
\caption{The Higgs potential and Electroweak symmetry breaking in THDM.}
\label{figHiggs}
\end{figure}

If we set $v_{1}=0$ in these equations, then we can find $v_{2}^{2}=\mu_{2}^{2}/\lambda_{2}^{2}$, and $\mu_{1}$ becomes a free parameter. Before proceeding, one needs to check whether the solution set with $v_{1}=0$ yields a stable minimum. Figure \ref{figHiggs} shows how the Higgs potential evolves with the VEVs of $\Phi_{1}$ and $\Phi_{2}$. We have fixed the free parameters to some values by using our numerical data, and as we can see, these values of the free parameters respect that the solutions with $v_{1}=0$ correspond to a stable minimum. All the electroweak data can be provided only with one VEV ($v_{2} \sim 200$ GeV), and as seen from the right panel of Figure \ref{figHiggs}, the potential looks exactly like that of the SM when the electroweak symmetry is broken.

The masses for the SM gauge bosons can be obtained through the kinetic term in the Lagrangian associated with the two scalar doublets. In general, these masses are functions of the two VEVs as $\propto v_{1}^{2}+v_{2}^{2}$ that requires, then, $v_{1}^{2}+v_{2}^{2} = v_{{\rm SM}}^{2}$. In our case, the model provides only one non-zero VEV ($v_{2}$), and hence, $v_{2}\simeq v_{{\rm SM}}$, and all the gauge boson masses have the same form as those in the SM. 

In addition to the gauge boson masses, the model yields five physical Higgs bosons in the particle spectrum. Diagonalizing the Higgs potential given in Eq.(\ref{HiggspotZ2}) their tree-level masses can be found as follows:

\begin{equation}
\setstretch{2.5}
\begin{array}{lll}
m_{h_{1}}^{2}=-\mu_{1}^{2}+\dfrac{1}{2}(\lambda_{3}+\lambda_{5})v_{2}^{2}~, & & m_{h_{2}}^{2}=2\lambda_{2}v_{2}^{2}, \\
m_{A}^{2}=-\mu_{1}^{2}+\dfrac{1}{2}(\lambda_{4}+\lambda_{5})v_{2}^{2}~, & & m_{H^{\pm}}^{2}=-\mu_{1}^{2}+\dfrac{1}{2}\lambda_{5}v_{2}^{2}
\end{array}
\label{Hmasses}
\end{equation}
where $m_{h_{1}}$ and $m_{h_{2}}$ are the masses of the CP-even Higgs boson mass eigenstates, while $m_{A}$ stands for the mass of the CP-odd Higgs boson, and $m_{H^{\pm}}$ denotes the mass of the charged Higgs bosons. Note that these masses depend on $\mu_{1}$ and the couplings $\lambda_{i}$, $i=2,3,4,5$. Since these terms are free parameters in our set up, the lightest CP-even Higgs boson state can be either $h_{1}$ or $h_{2}$ depending on the values of the relevant free parameters. It is seen from Eq(\ref{Hmasses}), $m_{A}^{2}-m_{H^{\pm}}^{2}=0.5 \lambda_{4}v_{2}^{2}$ at tree level. 

\begin{figure}[h!]
\centering
\subfigure{\includegraphics[scale=1]{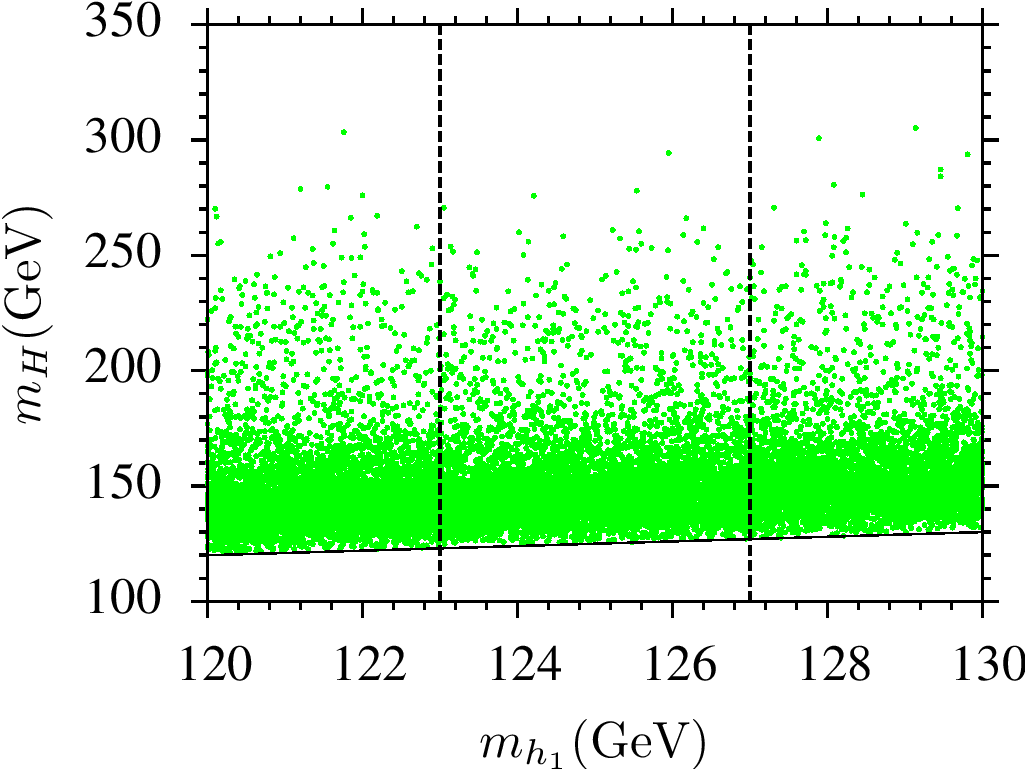}}
\subfigure{\includegraphics[scale=1]{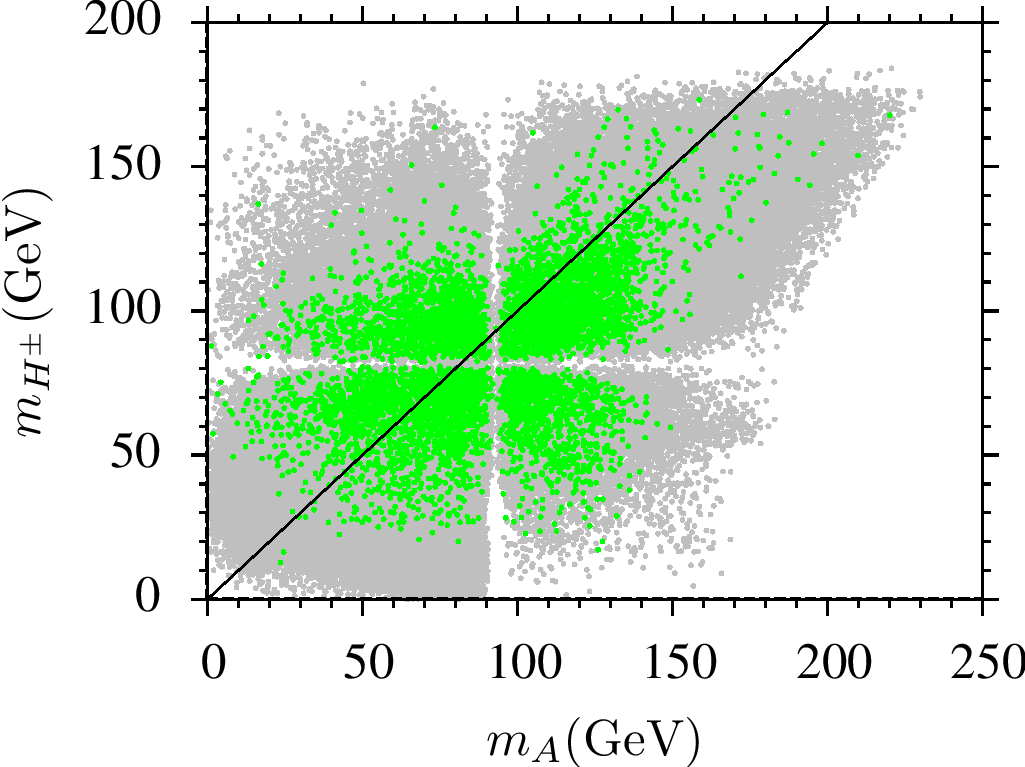}}
\subfigure{\includegraphics[scale=1]{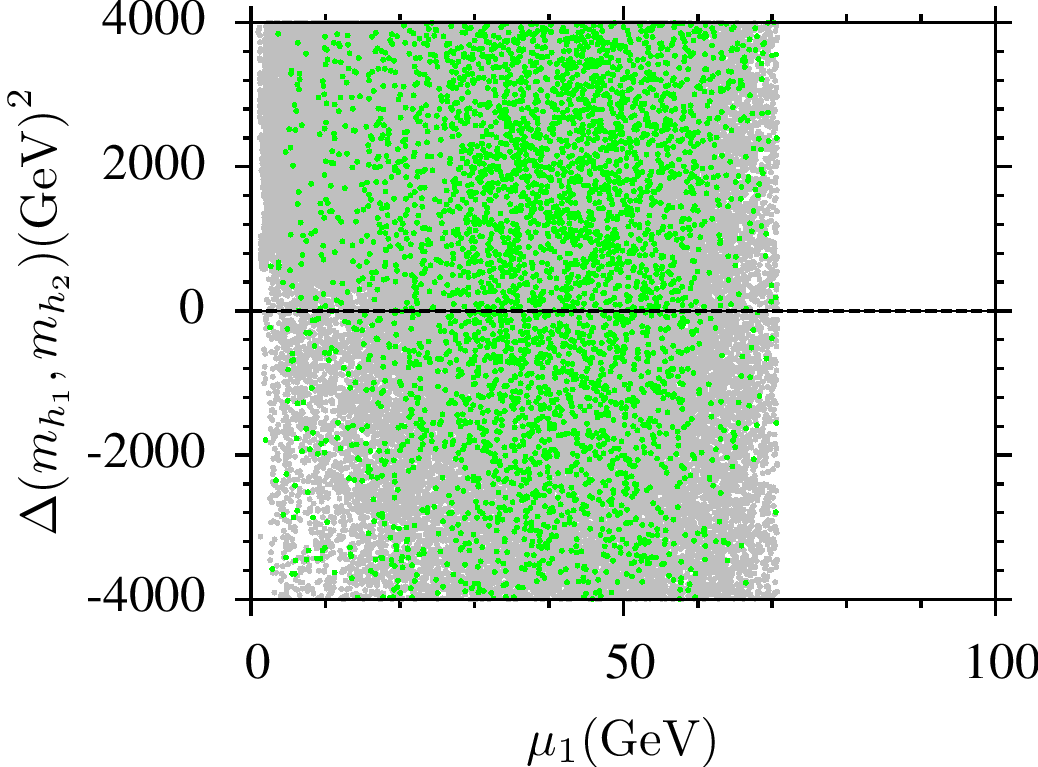}}
\subfigure{\includegraphics[scale=1]{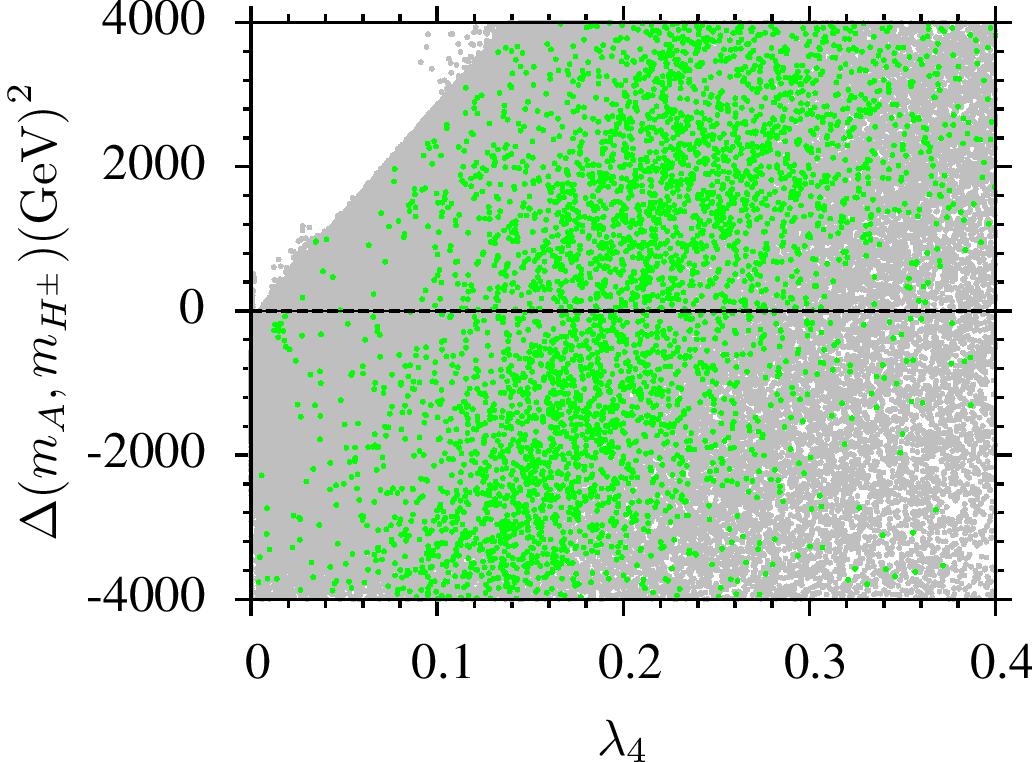}}
\caption{Plots in the $m_{H}-m_{h}$, $m_{H^{\pm}}-m_{A}$, $\Delta(m_{h_{1}},m_{h_{2}})-\mu_{1}$, and $\Delta(m_{A},m_{H^{\pm}})-\lambda_{4}$ planes. All points are consistent with the electroweak symmetry breaking, while green points also satisfy the experimental constraints discussed in Section \ref{sec:scan}, except the $m_{H}-m_{h}$ plane where the Higgs boson mass constraint is not applied, since the axes represent them directly. The vertical lines indicate the region $123 \leq m_{h} \leq 127$ GeV, and the diagonal lines indicate the degeneracy between the masses plotted.}
\label{figHmasses}
\end{figure}

Figure \ref{figHmasses} represents the Higgs boson masses with plots in  the $m_{H}-m_{h}$, $m_{H^{\pm}}-m_{A}$, $\Delta(m_{h_{1}},m_{h_{2}})-\mu_{1}$, and $\Delta(m_{A},m_{H^{\pm}})-\lambda_{4}$ planes. All points are consistent with the electroweak symmetry breaking, while green points also satisfy the experimental constraints discussed in Section \ref{sec:scan}, except the $m_{H}-m_{h}$ plane where the Higgs boson mass constraint is not applied, since the axes represent the Higgs boson masses directly. The vertical lines indicate the region $123 \leq m_{h} \leq 127$ GeV, and the diagonal lines indicate the degeneracy between the masses plotted. Since one of the Higgs doublets does not develop a non-zero VEV, the mass spectrum includes all the Higgs boson mass eigenstates with $m_{h,H,A,H^{\pm}} \lesssim 300$ GeV, as seen from the $m_{H}-m_{h}$ and $m_{H^{\pm}}-m_{A}$ planes. The $m_{H^{\pm}}-m_{A}$ plane also shows that $A$ and $H^{\pm}$ can be as light as a few GeV, while the solutions still yield the SM-like Higgs boson of mass about 125 GeV. A further exclusion limit on $m_{A}$ is usually applied as a function of $\tan\beta$ \cite{Khachatryan:2014wca}. Even though the solutions with $m_{A} \lesssim 200$ GeV can be excluded for $\tan\beta$ values, such exclusion limits are not well-defined when $\tan\beta=0$ as imposed when $v_{1}=0$ in our setup.

The $\Delta(m_{H},m_{h})-\mu_{1}$ plane display mass difference with $\Delta(m_{h_{1}},m_{h_{2}})\equiv m_{h_{1}}^{2}-m_{h_{2}}^{2}$ as a function of $\mu_{1}$. The lightest CP-even Higgs boson mass eigenstate is formed by $h_{1}$ when $\Delta(m_{h_{1}},m_{h_{2}}) < 0 $, while by $h_{2}$ when $\Delta(m_{h_{1}},m_{h_{2}}) > 0$. The horizontal line at $\Delta(m_{h_{1}},m_{h_{2}})=0$ corresponds to the solutions with $m_{h_{1}}=m_{h_{2}}$, which correspond to $m_{h}=m_{H}$ and hence imply two degenerate CP-even Higgs bosons in the spectrum. Meanwhile, despite being a free parameter, $\mu_{1}$ can be as large as only about 70 GeV to be consistent with the electroweak symmetry breaking. Similarly, we represent the mass difference between $A$ and $H^{\pm}$ bosons with $\Delta(m_{A},m_{H^{\pm}}) \equiv m_{A^{2}}-m_{H^{\pm}}^{2} $ as a function of $\lambda_{4}$ in the $\Delta(m_{A},m_{H^{\pm}})-\lambda_{4}$. Even though it would be expected to be a linear in $\lambda_{4}$, the loop corrections to the masses scatter the points, but increasing in the mass difference between these two bosons with $\lambda_{4}$ can be seen.

\section{ Comparison to the SM}
\label{sec:comparison}

As discussed in the previous section, the spectrum includes all the Higgs bosons below the mass scale at about 300 GeV. In addition, both $h_{1}$ and $h_{2}$ can play a role as the lightest CP-even Higgs boson, and when it is identified as the SM-like Higgs boson, they can change nature in how the SM-like Higgs boson couples to the fermions and gauge bosons, since $h_{1}$ and $h_{2}$ are realized as different superpositions of $\Phi_{1}$ and $\Phi_{2}$. Moreover, the CP-odd and charged Higgs bosons can be as light as a few GeV, while the lightest CP-even Higgs boson mass can be maintained at about 125 GeV. Such interesting results need to be analyzed and constrained further, since they are all in the detectable regime. In this section, we first analyze the CP-even Higgs bosons and discuss whether they can satisfy the properties exhibited by the SM-like Higgs boson. Then, we enlarge our discussion to the other Higgs boson mass eigenstates. 

\begin{figure}[t!]
\centering
\subfigure{\includegraphics[scale=1]{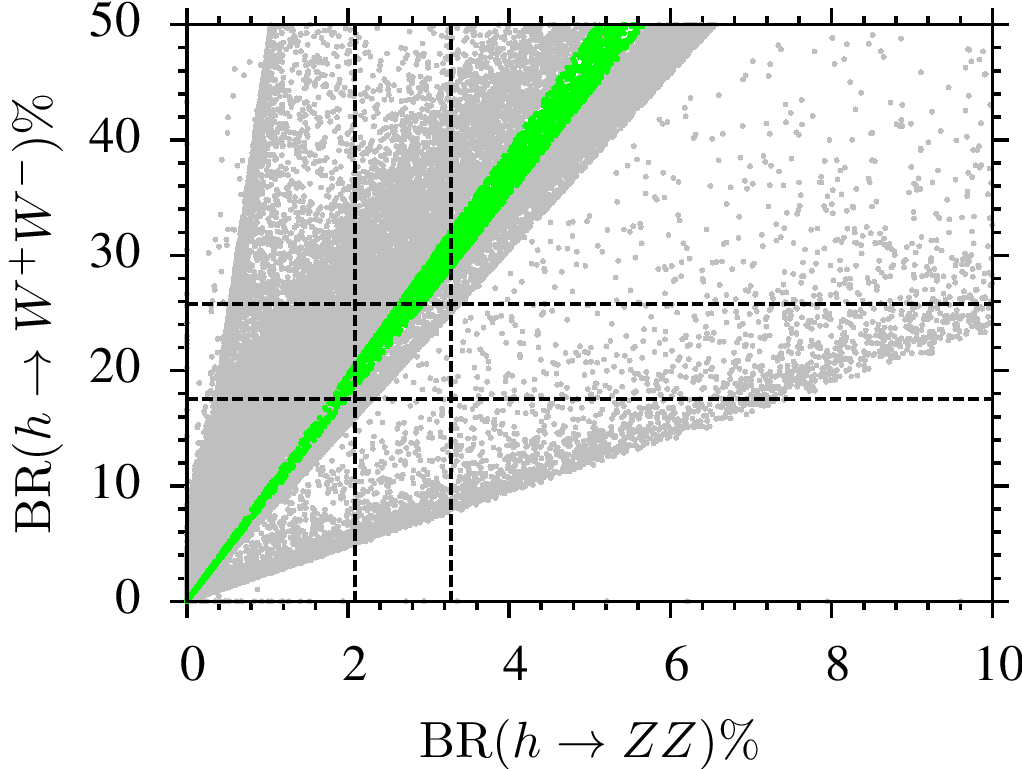}}
\subfigure{\includegraphics[scale=1]{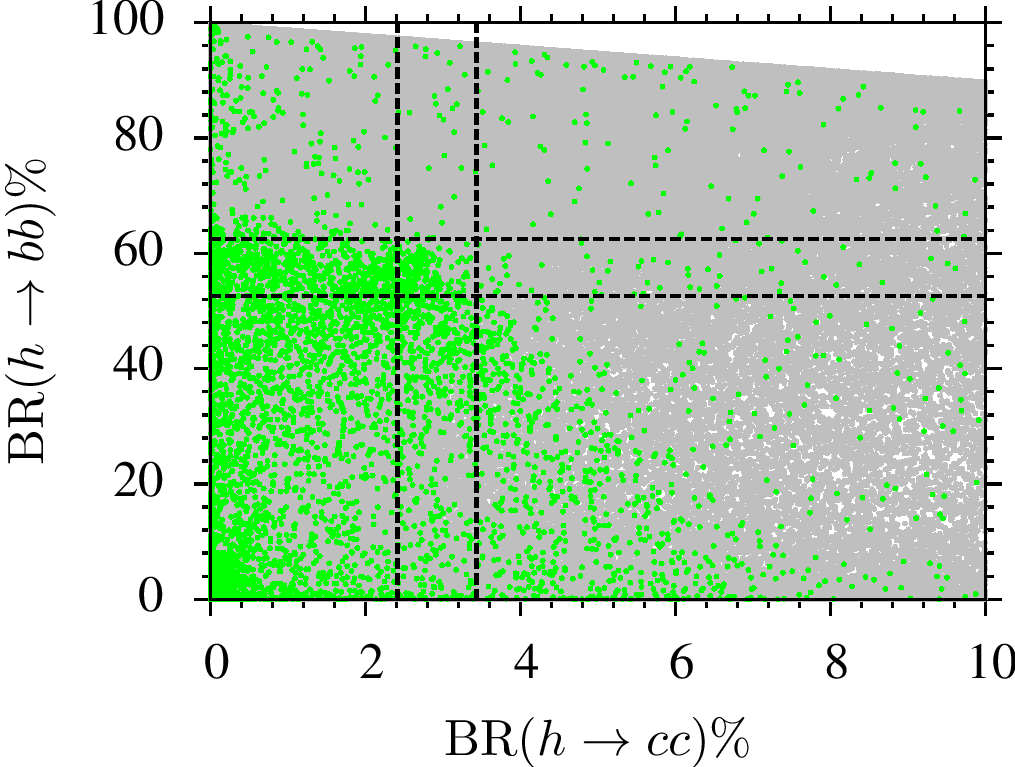}}
\subfigure{\includegraphics[scale=1]{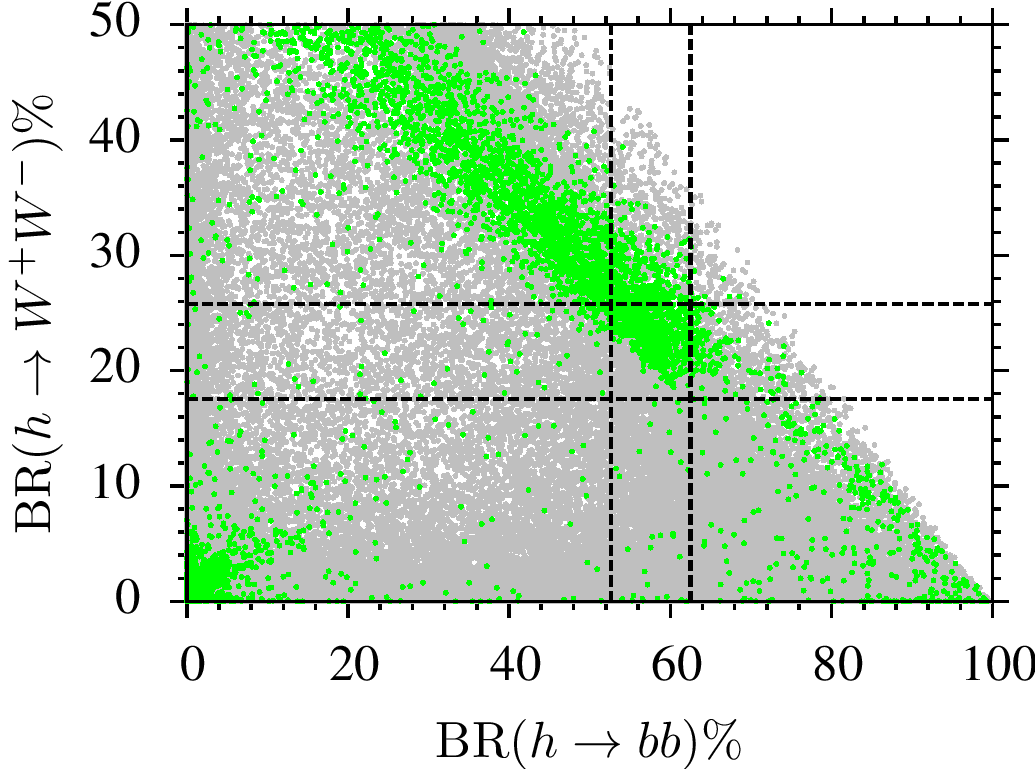}}
\subfigure{\includegraphics[scale=1]{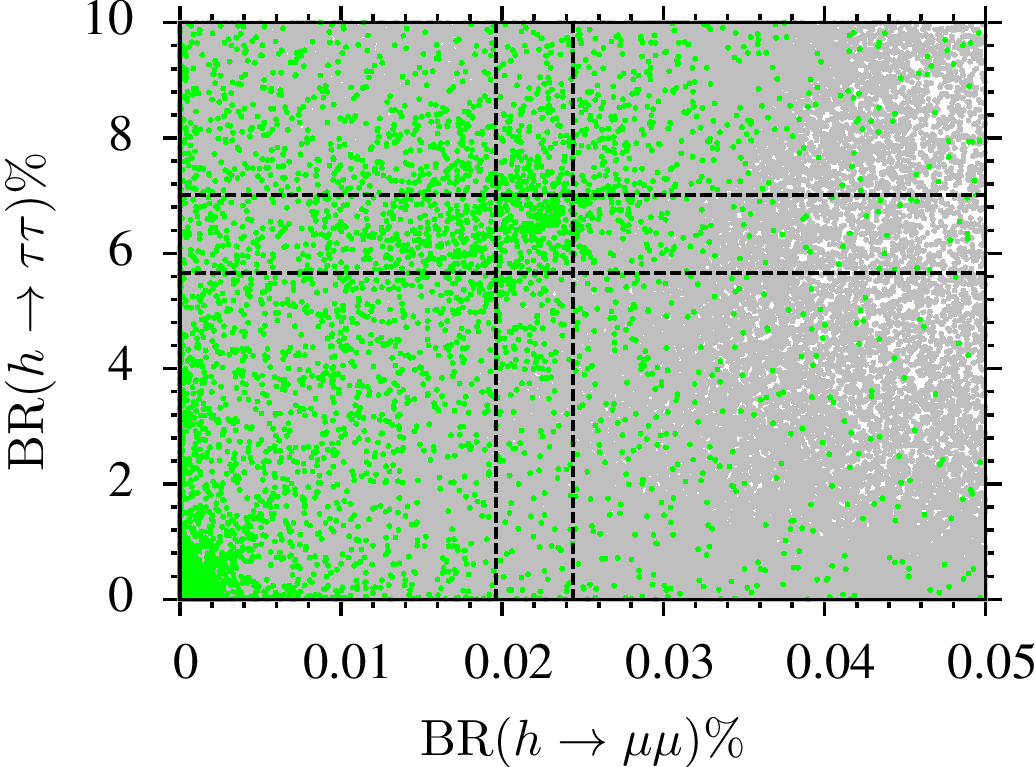}}
\caption{Results for the decay modes of $h$ with plots in the ${\rm BR}(h\rightarrow W+W-)-{\rm BR}(h\rightarrow ZZ)$, ${\rm BR}(h\rightarrow b\bar{b})-{\rm BR}(h\rightarrow c\bar{c})$, ${\rm BR}(h\rightarrow W+W-)-{\rm BR}(h\rightarrow b\bar{b})$, and ${\rm BR}(h\rightarrow \tau\bar{\tau})-{\rm BR}(h\rightarrow \mu\bar{\mu})$ planes. The color coding is the same as Figure \ref{figHmasses}. In addition, the dashed lines indicate the SM predictions \cite{Heinemeyer:2013tqa} for the plotted decay modes.}
\label{fig:hh1decays}
\end{figure}

\setcounter{footnote}{0}
Figure \ref{fig:hh1decays} displays our results for the decay modes of the lightest CP-even Higgs boson ($h$) with plots in the ${\rm BR}(h\rightarrow W+W-)-{\rm BR}(h\rightarrow ZZ)$, ${\rm BR}(h\rightarrow b\bar{b})-{\rm BR}(h\rightarrow c\bar{c})$, ${\rm BR}(h\rightarrow W^{+}W^{-})-{\rm BR}(h\rightarrow b\bar{b})$, and ${\rm BR}(h\rightarrow \tau\bar{\tau})-{\rm BR}(h\rightarrow \mu\bar{\mu})$ planes. The color coding is the same as Figure \ref{figHmasses}. In addition, the dashed lines indicate the SM predictions \cite{Heinemeyer:2013tqa} for the plotted decay modes. In the plots of Figure \ref{fig:hh1decays}, we analyze the results in the case when $h$ is assumed to be the one exhibiting properties of the SM's Higgs boson. Due to the uncertainties in calculation of the Higgs boson mass, mentioned in Section \ref{sec:scan}, we applied the SM predictions for the masses $m_{h}=123$ and $127$ GeV\footnote{The table for decay modes of the SM Higgs boson can be found at \href{https://twiki.cern.ch/twiki/bin/view/LHCPhysics/CERNYellowReportPageBR2014}{CERN Yellow Report Page}}. The most stringent bounds on the Higgs boson decay modes come from the channels $h\rightarrow W^{+}W^{-}$ and $h\rightarrow b\bar{b}$. The ${\rm BR}(h\rightarrow W^{+}W^{-})-{\rm BR}(h\rightarrow ZZ)$ plane shows that most of the solutions are excluded by the $h\rightarrow W^{+}W^{-}$ mode. Besides, it reveals a correlation with $h\rightarrow ZZ$ channel, and even though there is an observed excess in $h\rightarrow ZZ$ mode \cite{Chatrchyan:2013mxa}, this excess cannot be obtained without deviating from the SM prediction for  ${\rm BR}(h\rightarrow W^{+}W^{-})$. 

Similarly the $h\rightarrow b\bar{b}$ excludes most of the solutions, although a small region can still satisfy the SM predictions. The ${\rm BR}(h\rightarrow b\bar{b})-{\rm BR}(h\rightarrow c\bar{c})$ plane also indicates an inverse correlation with the $h\rightarrow c\bar{c}$ decay channel, although a few solutions can still yield large ${\rm BR}(h\rightarrow c\bar{c})$, while ${\rm BR}(h\rightarrow b\bar{b})$ remains within the SM prediction. The ${\rm BR}(h\rightarrow W^{+}W^{-})-{\rm BR}(h\rightarrow b\bar{b})$ plane compares the most strict channels, and the results show that satisfying one channel also leads to satisfy the other one. Finally, we display our results for the channels with two leptons in their final states in the ${\rm BR}(h\rightarrow \tau\bar{\tau})-{\rm BR}(h\rightarrow \mu\bar{\mu})$ plane. Even though the SM predictions again exclude most of the solutions, the uncertainties in these channels can cause to relax the bounds coming from these decay channels.

\begin{figure}[ht!]
\centering
\subfigure{\includegraphics[scale=1]{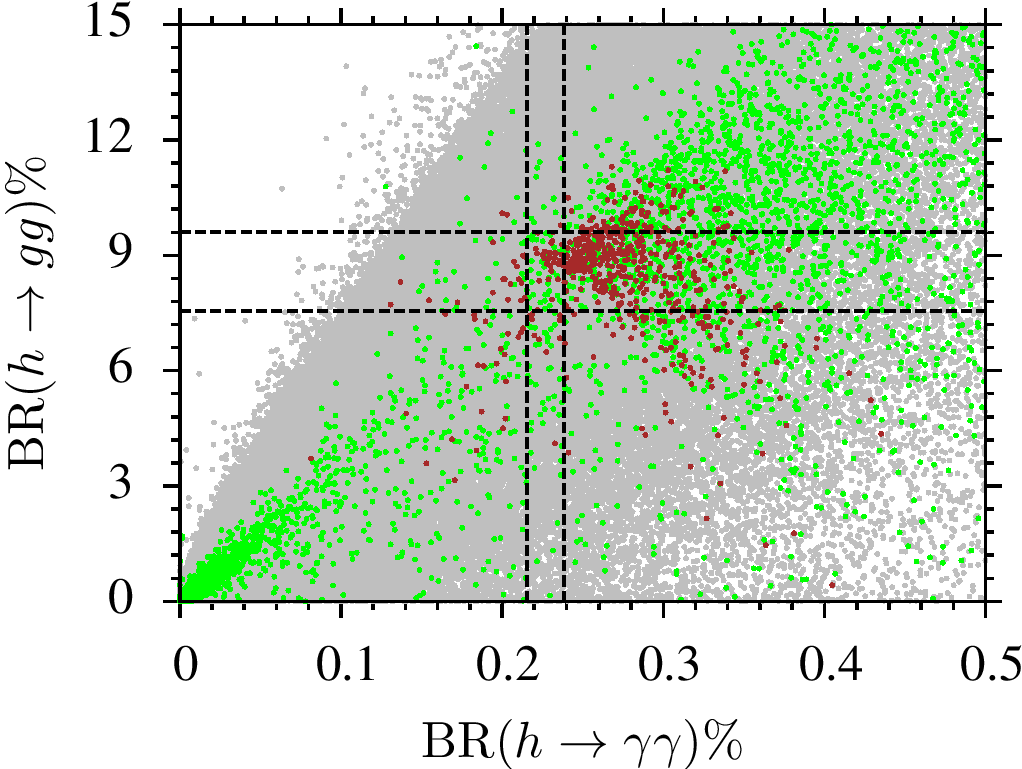}}
\subfigure{\includegraphics[scale=1]{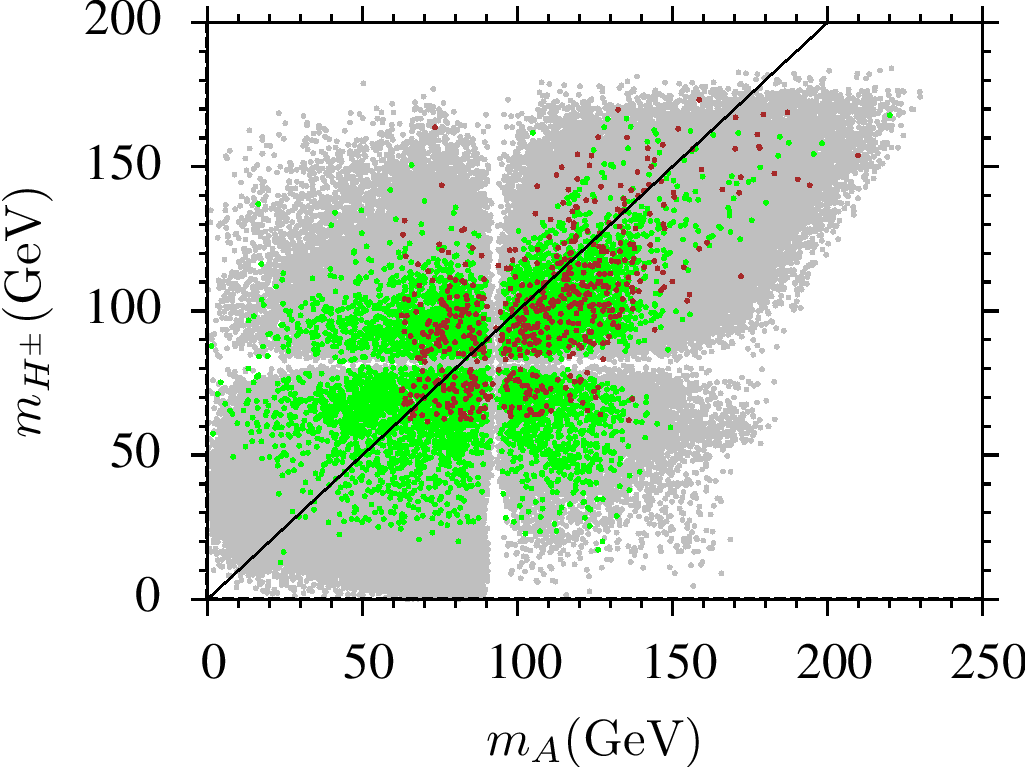}}
\caption{Plots in the ${\rm BR}(h\rightarrow gg)-{\rm BR}(h\rightarrow \gamma\gamma)$ and $m_{H^{\pm}}-m_{A}$ planes. The color coding is the same as Figure \ref{fig:hh1decays}. In addition, the brown points form a subset of green and they represent the solutions which satisfy the SM predictions for ${\rm BR}(h\rightarrow W^{+}W^{-})$ and ${\rm BR}(h\rightarrow b\bar{b})$.}
\label{fig:decayandmasses}
\end{figure}

Figure \ref{fig:decayandmasses} represents our results for the loop induced Higgs boson decays into a gluon and a photon pair, and the masses of the charged and CP-odd Higgs bosons with the plots in the ${\rm BR}(h\rightarrow gg)-{\rm BR}(h\rightarrow \gamma\gamma)$ and $m_{H^{\pm}}-m_{A}$ planes. The color coding is the same as Figure \ref{fig:hh1decays}. In addition, the brown points form a subset of green and they represent the solutions which satisfy the SM predictions for ${\rm BR}(h\rightarrow W^{+}W^{-})$ and ${\rm BR}(h\rightarrow b\bar{b})$. The ${\rm BR}(h\rightarrow gg)-{\rm BR}(h\rightarrow \gamma\gamma)$ shows that the excess in the $h\rightarrow \gamma \gamma$ channel \cite{CMS:ril} for the Higgs boson of mass about 125 GeV can be realized consistently with the other decay modes such as $h\rightarrow W^{+}W^{-},b\bar{b},gg$ (brown points between the horizontal dashed lines). As seen from the $m_{H^{\pm}}-m_{A}$ plane, the solutions with light charged and CP-odd Higgs bosons ($m_{A,H^{\pm}} \lesssim 60$ GeV) are excluded by the Higgs boson decays into $W^{+}W^{-}$ and $b\bar{b}$.

\begin{figure}[ht!]
\centering
\subfigure{\includegraphics[scale=1]{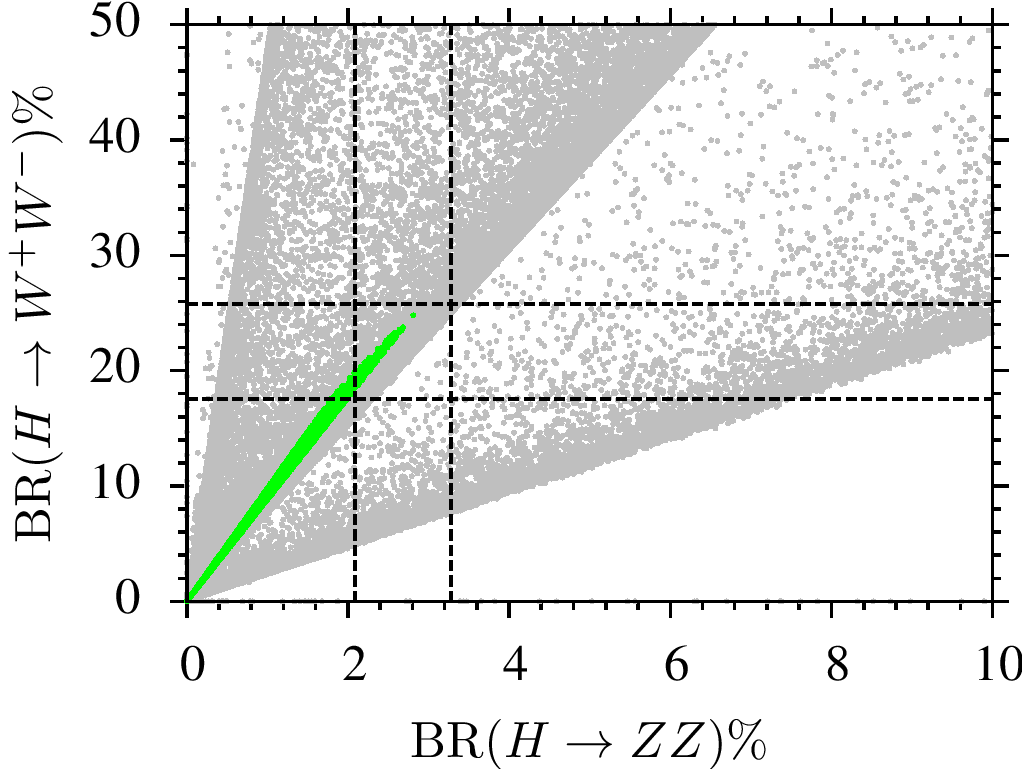}}
\subfigure{\includegraphics[scale=1]{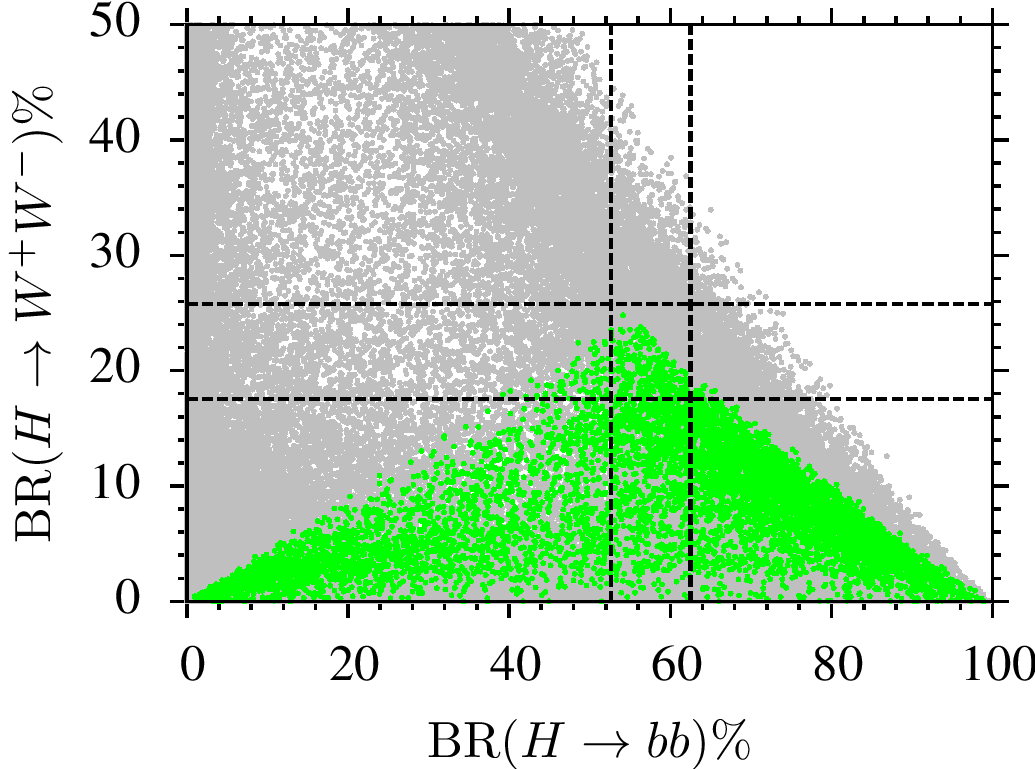}}
\subfigure{\includegraphics[scale=1]{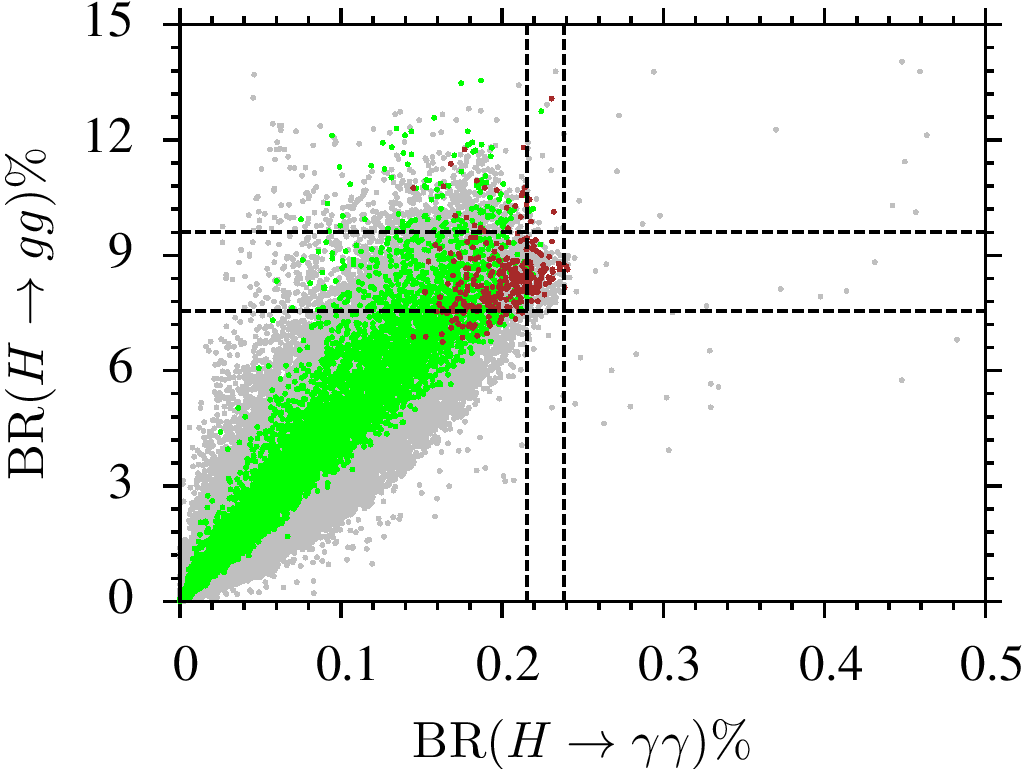}}
\subfigure{\includegraphics[scale=1]{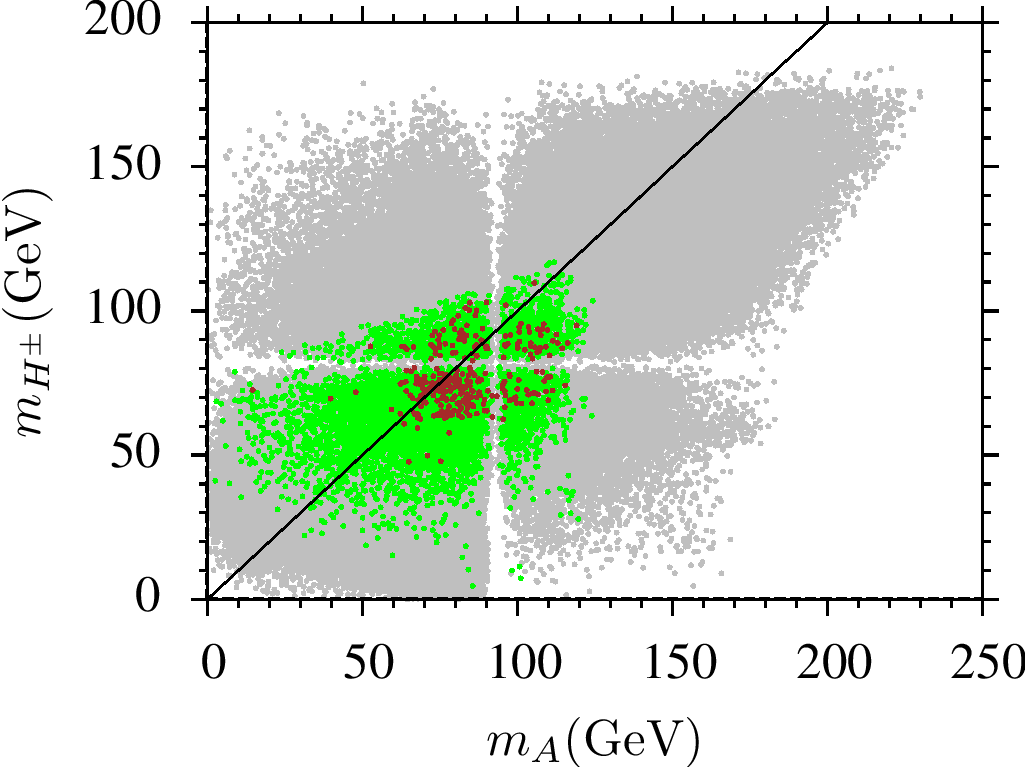}}
\caption{Results for the decay modes of $H$ with plots in the ${\rm BR}(H\rightarrow W+W-)-{\rm BR}(H\rightarrow ZZ)$, ${\rm BR}(H\rightarrow W^{+}W^{-})-{\rm BR}(H\rightarrow b\bar{b})$, ${\rm BR}(H\rightarrow gg)-{\rm BR}(h\rightarrow \gamma\gamma)$ and $m_{H^{\pm}}-m_{A}$ planes  The color coding is the same as Figure \ref{fig:hh1decays}, except we assume the heaviest CP-even Higgs boson (H) to be the SM-like one.}
\label{fig:hh2decays}
\end{figure}

We perform a similar analyses over the decay modes for the heaviest CP-even Higgs boson ($H$) in cases in which $H$ (instead of $h$) exhibits the SM-like Higgs boson properties. Our results are represented in Figure \ref{fig:hh2decays} with plots in the ${\rm BR}(H\rightarrow W+W-)-{\rm BR}(H\rightarrow ZZ)$, ${\rm BR}(H\rightarrow W^{+}W^{-})-{\rm BR}(H\rightarrow b\bar{b})$, ${\rm BR}(H\rightarrow gg)-{\rm BR}(h\rightarrow \gamma\gamma)$ and $m_{H^{\pm}}-m_{A}$ planes  The color coding is the same as Figure \ref{fig:hh1decays}. The linear correlation between the $W^{+}W^{-}$ and $ZZ$ decay modes are also realized for $H$, while the branching ratios for these channels cannot exceed the SM predictions. The ${\rm BR}(H\rightarrow W^{+}W^{-})-{\rm BR}(H\rightarrow b\bar{b})$ plane shows that the $W^{+}W^{-}$ and $b\bar{b}$ modes increase together up to ${\rm BR}(H\rightarrow W^{+}W^{-}) \sim 20\%$. Above this value, ${\rm BR}(H\rightarrow W^{+}W^{-})$ turns to decrease, while ${\rm BR}(H\rightarrow b\bar{b})$ keeps increasing. In contrast to the results for $h$, the excess in $H\rightarrow \gamma\gamma$ cannot be realized, as is seen in the ${\rm BR}(H\rightarrow gg)-{\rm BR}(h\rightarrow \gamma\gamma)$ plane. Finally, the $m_{H^{\pm}}-m_{A}$ plane shows that when the SM predictions are applied to $H$, the light masses for the CP-odd and charged Higgs boson solutions ($m_{A},m_{H^{\pm}} \lesssim 50$ GeV) are excluded. After all, even though the exclusion limit on $A$ and $H^{\pm}$ cannot be applied properly when $\tan\beta =0$, their mass scales can be constrained from below by the SM predictions for the Higgs sector.

\section{Higgs Boson Production at LHC}

\begin{figure}[ht!]
\centering
\subfigure{\includegraphics[scale=1]{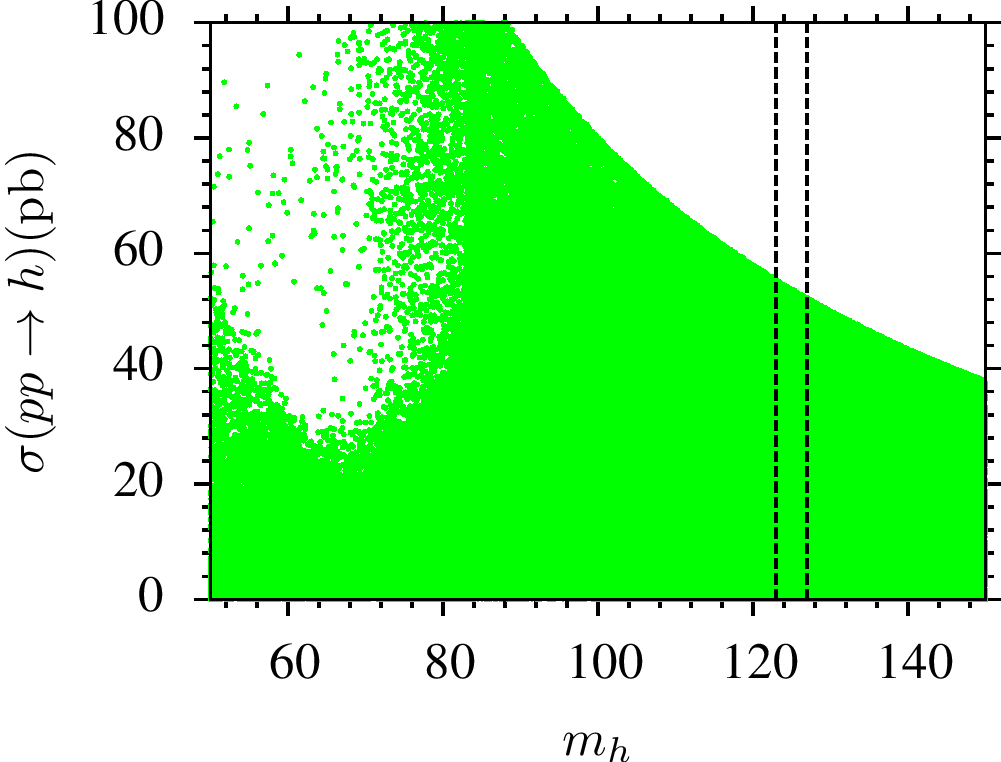}}
\subfigure{\includegraphics[scale=1]{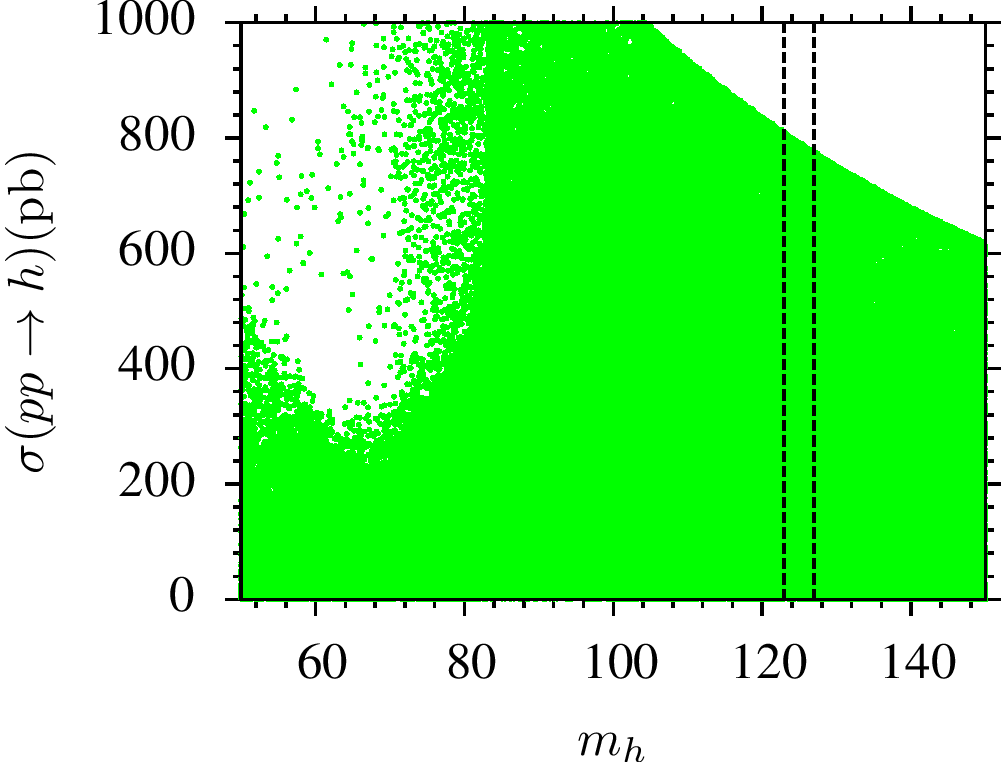}}
\caption{Higgs boson production at colliders with 14 TeV (left) and 100 TeV (right). The color coding is the same as Figure \ref{figHmasses} except the Higgs mass bound is not applied in the gren region to analyze the production rates for all possible mass scales of the Higgs boson.}
\label{fig:higgsprod}
\end{figure}

The results considered in the previous section are rather based on the decay channels, and even though they can give some hints, if they can provide a possible signal, the production processes for the Higgs bosons should be employed as well as the branching ratios for the relevant decay modes. Figure \ref{fig:higgsprod} displays the production cross-sections for the Higgs boson at the colliders with 14 TeV (left) and 100 TeV (right) center of mass (COM) energies. The color coding is the same as Figure \ref{figHmasses} except the Higgs mass bound is not applied in the green region to analyze the production rates for all possible mass scales of the Higgs boson. The current experiments are being conducted at 14 TeV, while the LHC is planned to reach to the COM energy of 100 TeV in future. The Higgs boson is produced mostly through the gluon fusion, while the other possible channels can be considered only some corrections in comparison to the Higgs boson production cross-section at the LHC. The production cross-section is about 100 pb for the SM Higgs boson \cite{Djouadi:2005gj}, while it is about 60 pb in our setup for the THDM, as seen from the left panel of Figure \ref{fig:higgsprod}. These results yield a reduction of about $40\%$ in the cross-section of a possible signal process, and hence the branching ratios for the relevant decay modes, deviating from the SM predictions within such reduction rates, can be still allowed. 

In contrast to the experiments with 14 TeV COM energy, at the Future Circular Collider (FCC) experiments, whose reach is planned to be 100 TeV COM energy, the Higgs production with mass about 125 GeV is expected to be about 800 pb, which coincides with the production cross-section of the SM Higgs boson \cite{Baglio:2015wcg}. Since the THDM framework with the $Z_{2}$ symmetry predicts more or less the same prediction rate for the Higgs boson, it yields quite similar implications to the SM predictions for the Higgs boson, and the experimental results can be covered within THDM.

If we assume $H$ is the SM-like Higgs boson with mass of 125 GeV, then $h$ can be even as light as 50 GeV. Even though such a light scalar boson can escape from the detection at the current experiments, due to the small production rate in comparison to the SM Higgs boson production, FCC can provide efficient circumstances for a possible signal from such light scalar particle states, which can decay into the SM particles.

\begin{figure}[ht!]
\centering
\subfigure{\includegraphics[scale=1]{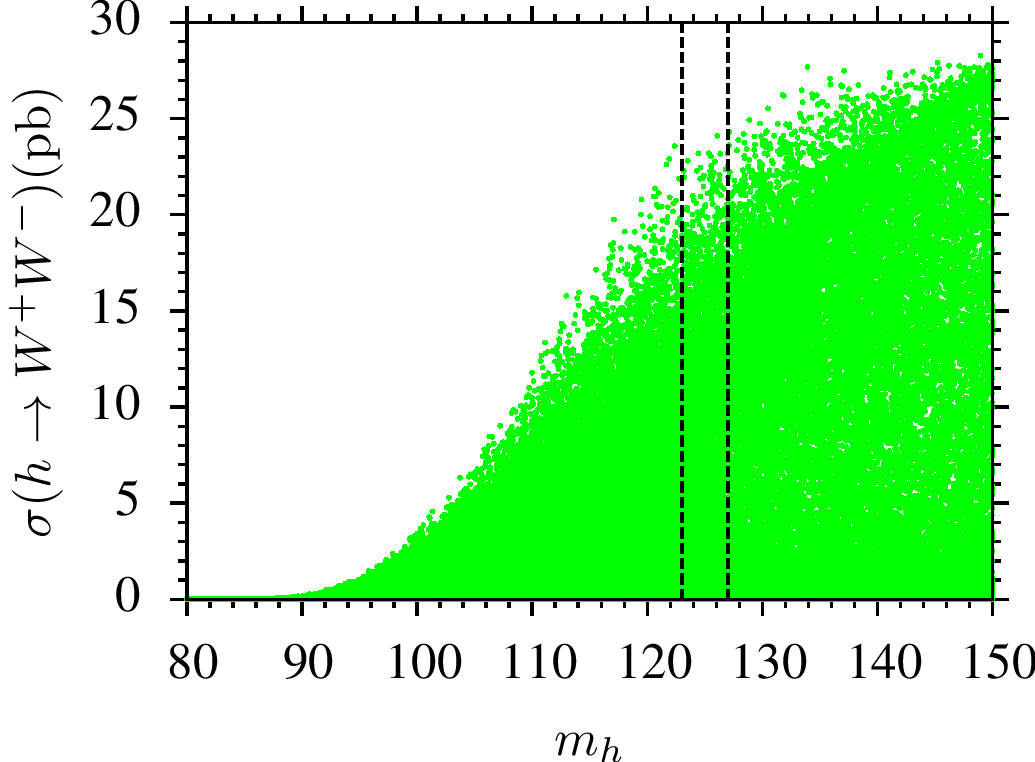}}
\subfigure{\includegraphics[scale=1]{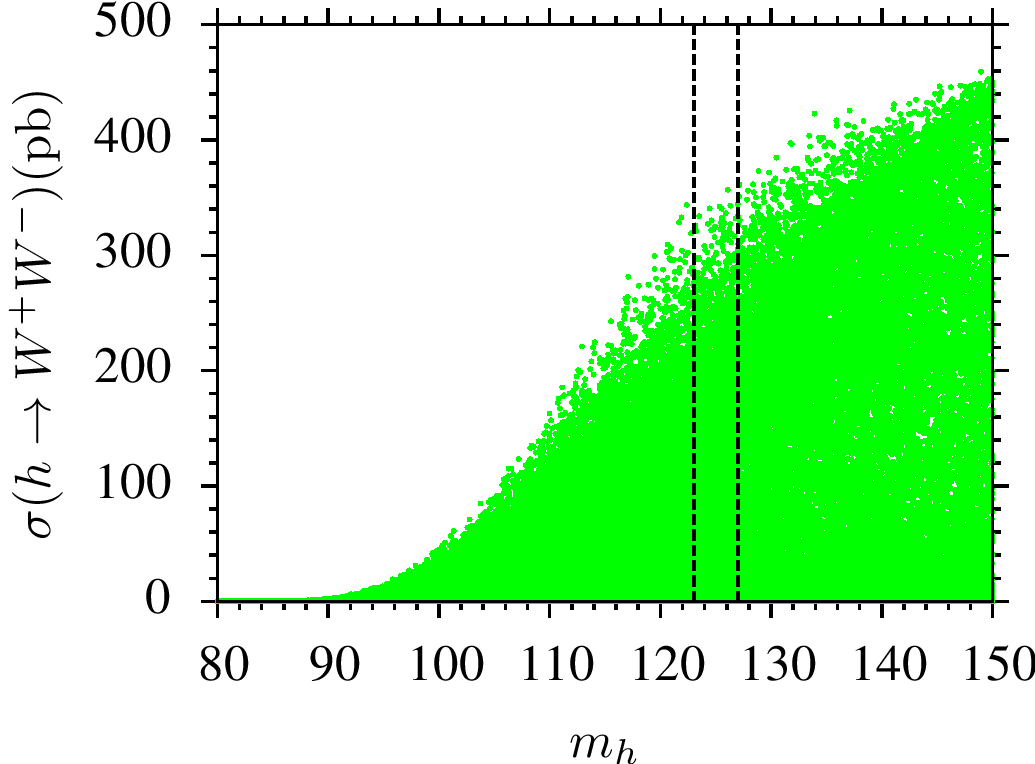}}
\subfigure{\includegraphics[scale=1]{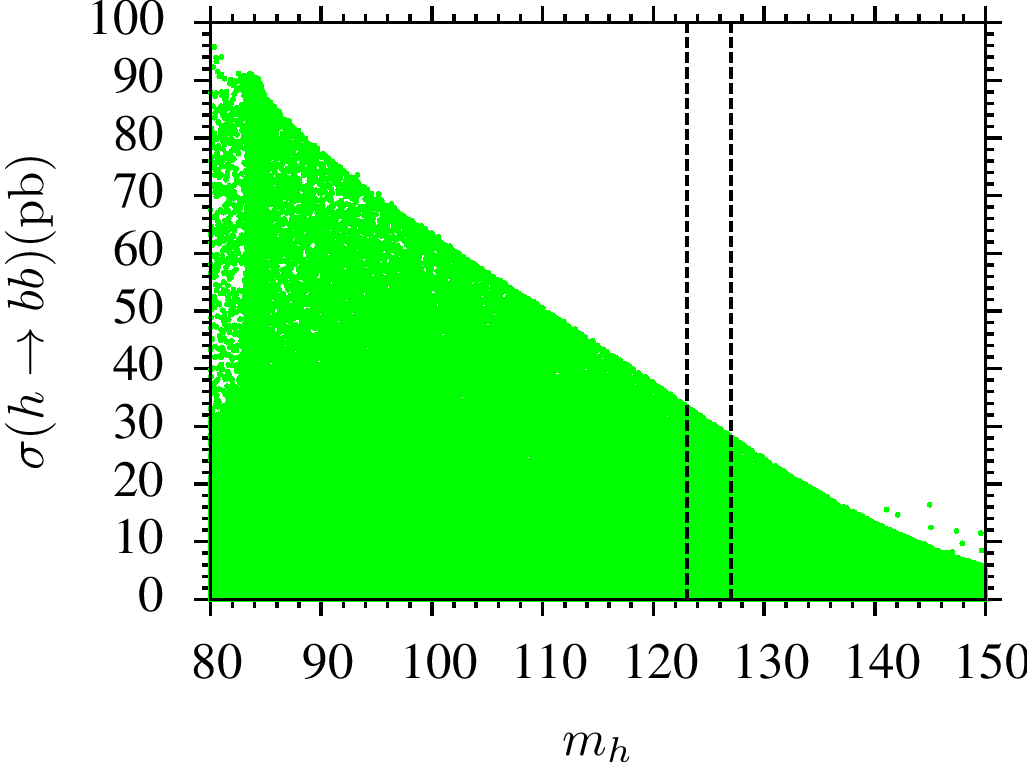}}
\subfigure{\includegraphics[scale=1]{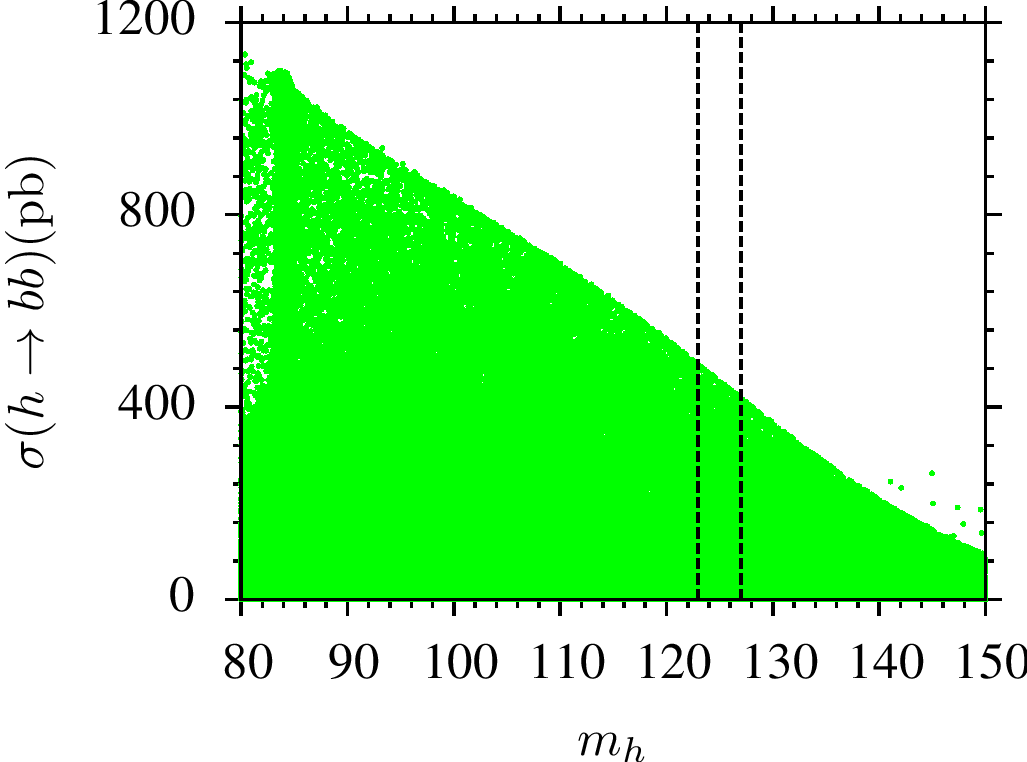}}
\caption{Cross-sections for the decay channels $h\rightarrow W^{+}W^{-}$ (top) and $h\rightarrow b\bar{b}$ (bottom) for the 14 TeV (left) and FCC (right) in a correlation with the Higgs boson mass $m_{h}$. The color coding is the same as Figure \ref{fig:higgsprod}.}
\label{figdecaycross}
\end{figure}

If we consider the cross-sections for the relevant decay modes, they can be calculated in a good approximation as follows:

\begin{equation}
\sigma(h\rightarrow XX) \simeq \sigma(pp\rightarrow h)\times {\rm BR}(h\rightarrow XX)~,
\label{XXcross}
\end{equation}
where $X$ represents possible SM particles. 

If we consider the Higgs boson decay modes by employing Eq.(\ref{XXcross}), we obtain the results shown in Figure \ref{figdecaycross} for the decay channels $h\rightarrow W^{+}W^{-}$ (top) and $h\rightarrow b\bar{b}$ (bottom) for the 14 TeV (left) and FCC (right) in a correlation with the Higgs boson mass $m_{h}$. The color coding is the same as Figure \ref{fig:higgsprod}. While the cross-section of $h\rightarrow W^{+}W^{-}$ increases with the Higgs boson mass, it decreases for $h\rightarrow b\bar{b}$ with increasing Higgs boson mass. The cross-section for the $W^{+}W^{-}$ decay mode is found at most $\sigma(h\rightarrow W^{+}W^{-}) \sim 25$ pb for the experiments with 14 TeV COM energy. Even though it can be considered high enough to be detected, the cross-section ($\sim 65$ pb without the SM Higgs boson) of the other decays with $W^{+}W^{-}$ final states in the SM is twice as large as what is shown in the top left panel of Figure \ref{figdecaycross}. The similar discussion can be followed for the $h\rightarrow b\bar{b}$, whose cross-section is about 40 pb with $m_{h}\sim 125$ GeV. The other SM processes ending with $b\bar{b}$ final states yield much larger cross-section ($\sim 10^{8}$ pb). Such processes with large cross-sections cause a strong background, which extremely suppress the signal. When the LHC is upgraded to FCC, the signal cross-sections are expected to reach $10-15$ times greater levels, but it will be true for the background processes as well.  

\section{Note on Muon $g-2$}

As discussed in Section \ref{sec:scan}, the extra Higgs boson can easily spoil the SM predictions for ${\rm BR}(b\rightarrow s\gamma)$. In order to avoid this conflict, either the Higgs bosons should be so heavy that the extra contributions to this process would be suppressed by their masses, or the relevant couplings should be small enough. On the other hand, the masses and couplings of the extra Higgs bosons are also important to resolve the muon $g-2$ discrepancy between the experiment and the SM. As shown above, the solutions significantly contributing to muon $g-2$ also leads to results for ${\rm BR}(b\rightarrow s\gamma)$ out of the SM predictions. Throughout our analyses, we do not insist on resolution to the muon $g-2$ resolution, but we require the solutions to be compatible with the SM prediction in ${\rm BR}(b\rightarrow s\gamma)$. 

\begin{figure}[ht!]
\centering
\subfigure{\includegraphics[scale=1]{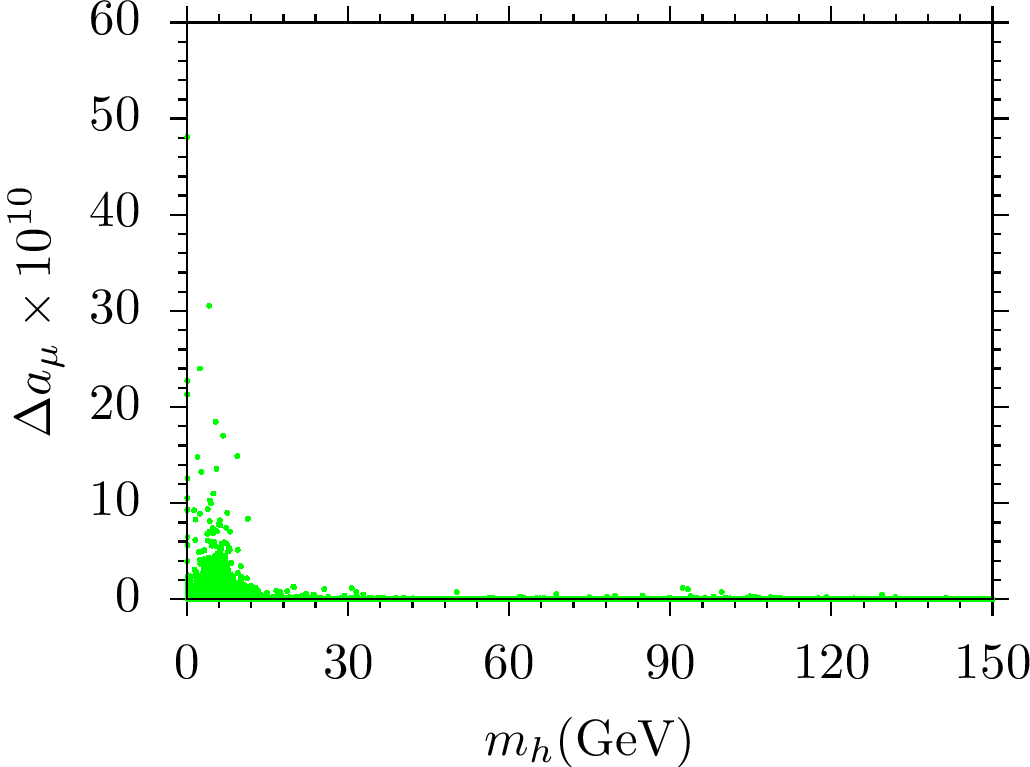}}
\subfigure{\includegraphics[scale=1]{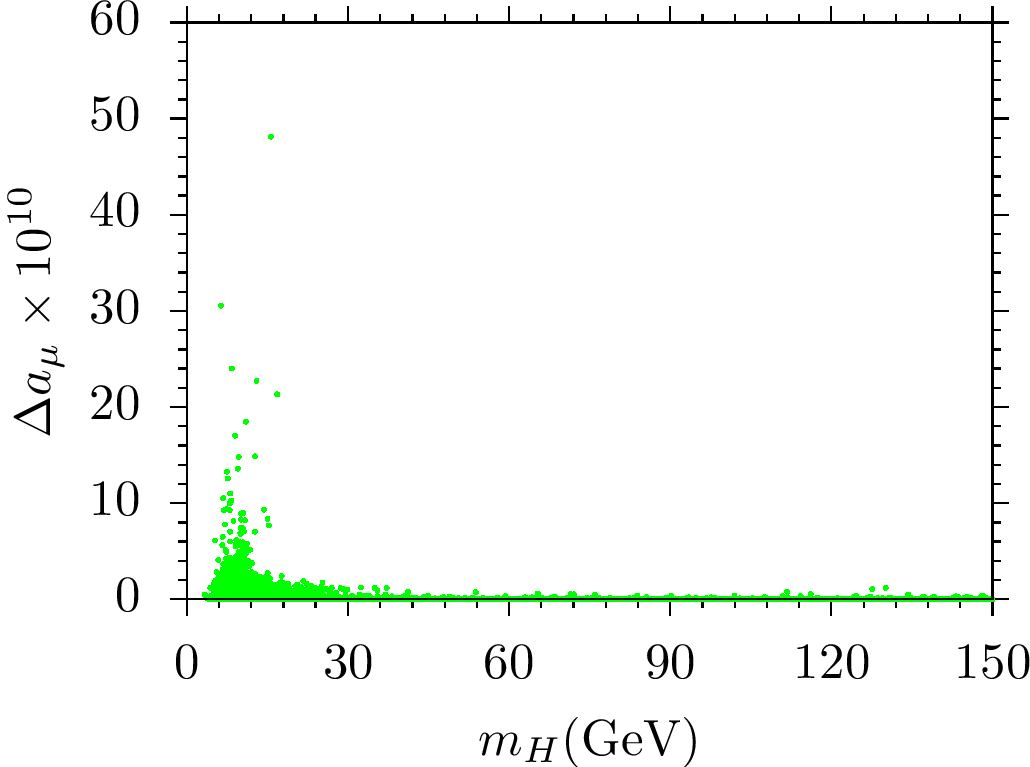}}
\caption{Muon $g-2$ results in correlation with the Higgs boson masses $m_{h}$ (left) and $m_{H}$ right. All points are consistent with EWSB. The green points satisfy only the constraint from ${\rm BR}(b\rightarrow s\gamma)$.}
\label{muong2}
\end{figure}

The discrepancy between the experiment \cite{Bennett:2006fi} and the SM \cite{Davier:2010nc} can be stated as 

\begin{equation}
\Delta a_{\mu}\equiv a_{\mu}^{{\rm exp}}-a_{\mu}^{{\rm SM}} =(28.7 \pm 8)\times 10^{-10}
\label{damuexp}
\end{equation}

If the model can provide enough contributions that can cover the experimental measurements, then the solutions can be considered as a sort of resolution to the muon $g-2$ problem. On the other hand, as is expected, such contributions are quite negligible ($\lesssim 2\times 10^{-10}$) as shown in Figure \ref{muong2} in correlation with the Higgs boson masses $m_{h}$ (left) and $m_{H}$ right. All points are consistent with EWSB. The green points satisfy only the constraint from ${\rm BR}(b\rightarrow s\gamma)$. Moreover, the region with the largest contributions to the muon $g-2$ cannot be accommodate with a 125 GeV Higgs boson, since both $m_{h}$ and $m_{H}$ are below 100 GeV in these regions. As a result, even though THDM can be considered an effective theory, which a larger model can be projected into after decoupling regimes, the muon $g-2$ resolution still needs to receive contributions from new particles other than the Higgs bosons.

\section{Conclusion}
\label{sec:conc}

We review the THDM model in a general fashion and focus on the cases in which one of the Higgs doublets does not develop a non-zero VEV (chosen as $v_{1}=0$). In this case, the Higgs fields with zero VEV does not contribute to the physical masses of the SM particles, and hence, its interactions with the SM particles can have more freedom than in the case of usual considerations on the THDM models. Such solutions cannot be considered in Type-II, since fermions acquire their masses from different Higgs doublets. In this sense, we set up the Yukawa Lagrangian such that each fermion interacts with both of the Higgs doublets (i.e. Type-III). In addition, Type-I can be identified as a submodel of Type-III, which can be obtained when the interaction couplings of a Higgs doublet are set to zero. We show that the stability of the Higgs potential minima can be maintained, even if one of the Higgs doublets does not have a non-zero VEV. Although such a freedom in interactions of the Higgs doublets can be seen as an advantage in the model, the strong agreement between the experiments and the SM predictions provides very stringent constraints. We find that all the Higgs bosons are lighter than about 300 GeV in the low scale spectrum when $v_{1}=0$. Such light mass scales are in the detectable regime, and hence, the SM predictions and the experimental results are essential to be applied in the analyses. Especially the decay channels; $h\rightarrow W^{+}W^{-}$ and $h\rightarrow b\bar{b}$ exclude most of the solutions, while it is still possible to realize a small region, which coincides with the SM predictions. We highlight that it is also possible to realize an excess in $h\rightarrow \gamma\gamma$ decay channel, even one applies the constraints from the Higgs boson decays into $W^{+}W^{-}$ and $b\bar{b}$. In addition, if one assumes $H$ is the SM-like Higgs boson of mass about 125 GeV, the solutions with $m_{h}\lesssim 125$ GeV can be acceptable. In this case, the solutions can still be realized consistent with the SM predictions;however, an excess in $H\rightarrow \gamma\gamma$ cannot be observed, while the implications for this channel in THDM can stay in the SM prediction rates at most.

One of the most important constraint on the Higgs sector comes from the flavor chancing rare $B-$meson decays, which can be identified with the $b\rightarrow s\gamma$ channel. The SM prediction on such decay processes are in a very good agreement with the experimental results, and hence even a small contribution from new particles can spoil the experimental confirmation. In THDM, especially the charged Higgs boson can significantly contribute to such processes, and hence, they bring a stringent constraint on the Higgs sector. In order to satisfy this constraint, either the Higgs bosons should be heavier than about 200 GeV, or they should negligibly couple to the SM particles. Despite the light Higgs bosons in the spectrum ($m_{H^{\pm}} \lesssim 200$ GeV), the solutions can still be consistent with the ${BR}(b\rightarrow s\gamma)$ measurements. When we apply this constrain as an essential ingredient in our analyses, its negative results can be identified in the muon $g-2$ implications. The disagreement between the experiment and the SM in muon $g-2$ predictions can be resolved by the contributions from the new particles, and such a resolution requires these new particles to be rather light. Even though the mass scales of the Higgs bosons are favored by muon $g-2$, such Higgs bosons cannot couple to the SM fermions (say muons in muon $g-2$ calculation) more than the SM to yield consistent $b\rightarrow s\gamma$. We show that it is not possible to accommodate the muon $g-2$ resolution with the consistent $B-$meson decays. 

{\bf Acknowledgments}

\noindent
We would like to thank Zafer Alt\i n, Volkan Aslan for their useful discussions and comments. This work is supported in part by The Scientific and Technological Research Council of Turkey (TUBITAK) Grant no. MFAG-114F461 (CS\"{U}, A\c{C}). Part of the numerical calculations reported in this paper were performed at the National Academic Network and Information Center (ULAKBIM) of TUBITAK, High Performance and Grid Computing Center (TRUBA resources).


\begin{thebibliography}{99}

\bibitem{Aad:2012tfa} 
  G.~Aad {\it et al.} [ATLAS Collaboration],
  Phys.\ Lett.\ B {\bf 716}, 1 (2012)
  doi:10.1016/j.physletb.2012.08.020
  [arXiv:1207.7214 [hep-ex]].
  
\bibitem{Chatrchyan:2012xdj} 
  S.~Chatrchyan {\it et al.} [CMS Collaboration],
  Phys.\ Lett.\ B {\bf 716}, 30 (2012)
  doi:10.1016/j.physletb.2012.08.021
  [arXiv:1207.7235 [hep-ex]].

\bibitem{CMS:ril} 
  [CMS Collaboration],
  CMS-PAS-HIG-13-001.

\bibitem{Chatrchyan:2013mxa} 
  S.~Chatrchyan {\it et al.} [CMS Collaboration],
  Phys.\ Rev.\ D {\bf 89}, no. 9, 092007 (2014)
  doi:10.1103/PhysRevD.89.092007
  [arXiv:1312.5353 [hep-ex]].

\bibitem{Gildener:1976ai}
  E.~Gildener,
  Phys.\ Rev.\ D {\bf 14}, 1667 (1976);
  E.~Gildener,
  Phys.\ Lett.\ B {\bf 92}, 111 (1980);
  S.~Weinberg,
  Phys.\ Lett.\ B {\bf 82}, 387 (1979);
L.~Susskind,
  Phys.\ Rev.\ D {\bf 20}, 2619 (1979);
  M.~J.~G.~Veltman,
  Acta Phys.\ Polon.\ B {\bf 12}, 437 (1981).


\bibitem{hinstability}
G.~Degrassi, S.~Di Vita, J.~Elias-Miro, J.~R.~Espinosa, G.~F.~Giudice, G.~Isidori and A.~Strumia,
  JHEP {\bf 1208}, 098 (2012)
  [arXiv:1205.6497 [hep-ph]];
  F.~Bezrukov, M.~Y.~.Kalmykov, B.~A.~Kniehl and M.~Shaposhnikov,
  JHEP {\bf 1210}, 140 (2012)
  [arXiv:1205.2893 [hep-ph]];
D.~Buttazzo, G.~Degrassi, P.~P.~Giardino, G.~F.~Giudice, F.~Sala, A.~Salvio and A.~Strumia,
  JHEP {\bf 1312}, 089 (2013)
  [arXiv:1307.3536].

\bibitem{Baer:2006rs} 
  H.~Baer and X.~Tata,
  ``Weak scale supersymmetry: From superfields to scattering events,'', Cambridge, Univ. Pr. (2006) 537p.

\bibitem{Branco:2011iw} 
  G.~C.~Branco, P.~M.~Ferreira, L.~Lavoura, M.~N.~Rebelo, M.~Sher and J.~P.~Silva,
  Phys.\ Rept.\  {\bf 516}, 1 (2012)
  [arXiv:1106.0034 [hep-ph]];
  R.~A.~Diaz,
  hep-ph/0212237;
  I.~Chakraborty and A.~Kundu,
  Phys.\ Rev.\ D {\bf 92}, no. 9, 095023 (2015)
  [arXiv:1508.00702 [hep-ph]];
  F.~Mahmoudi and O.~Stal,
  Phys.\ Rev.\ D {\bf 81}, 035016 (2010)
  [arXiv:0907.1791 [hep-ph]];
  I.~P.~Ivanov and J.~P.~Silva,
  Phys.\ Rev.\ D {\bf 92}, no. 5, 055017 (2015)
  [arXiv:1507.05100 [hep-ph]].

\bibitem{Aguilar-Saavedra:2013qpa} For an incomplete list, see \\
  J.~A.~Aguilar-Saavedra, R.~Benbrik, S.~Heinemeyer and M.~Pérez-Victoria,
  Phys.\ Rev.\ D {\bf 88}, no. 9, 094010 (2013)
  [arXiv:1306.0572 [hep-ph]];
%
  M.~Badziak,
  Phys.\ Lett.\ B {\bf 759}, 464 (2016)
  [arXiv:1512.07497 [hep-ph]];
%
  A.~Angelescu, A.~Djouadi and G.~Moreau,
  Phys.\ Lett.\ B {\bf 756}, 126 (2016)
  [arXiv:1512.04921 [hep-ph]];
%
 M.~Carena, I.~Low, N.~R.~Shah and C.~E.~M.~Wagner,
 JHEP {\bf 1404}, 015 (2014)
 [arXiv:1310.2248 [hep-ph]];
  J.~F.~Gunion, H.~E.~Haber, G.~L.~Kane and S.~Dawson,
  Front.\ Phys.\  {\bf 80}, 1 (2000);
 S.~A.~R.~Ellis, R.~M.~Godbole, S.~Gopalakrishna and J.~D.~Wells,
 JHEP {\bf 1409}, 130 (2014)
 [arXiv:1404.4398 [hep-ph]];
A.~Arhrib, R.~Benbrik, S.~J.~D.~King, B.~Manaut, S.~Moretti and C.~S.~Un,
  arXiv:1607.08517 [hep-ph].

\bibitem{Un:2016hji} 
  C.~S.~Un and O.~Ozdal,
  Phys.\ Rev.\ D {\bf 93}, 055024 (2016)
  doi:10.1103/PhysRevD.93.055024
  [arXiv:1601.02494 [hep-ph]], and references therein.

\bibitem{Porod:2003um} 
  W.~Porod,
  Comput.\ Phys.\ Commun.\  {\bf 153}, 275 (2003)
  doi:10.1016/S0010-4655(03)00222-4
  [hep-ph/0301101].

\bibitem{Staub:2008uz} 
  F.~Staub,
  arXiv:0806.0538 [hep-ph];
  F.~Staub,
  Comput.\ Phys.\ Commun.\  {\bf 182}, 808 (2011)
  [arXiv:1002.0840 [hep-ph]];
  F.~Staub,
  Comput.\ Phys.\ Commun.\  {\bf 185}, 1773 (2014)
  [arXiv:1309.7223 [hep-ph]].

\bibitem{Belanger:2009ti}
  G.~Belanger, F.~Boudjema, A.~Pukhov and R.~K.~Singh,
  JHEP {\bf 0911}, 026 (2009);
H.~Baer, S.~Kraml, S.~Sekmen and H.~Summy,
  JHEP {\bf 0803}, 056 (2008).

\bibitem{Group:2009ad} 
  T.~E.~W.~Group [CDF and D0 Collaborations],
  arXiv:0903.2503 [hep-ex].

\bibitem{Gogoladze:2011aa} 
  I.~Gogoladze, Q.~Shafi and C.~S.~Un,
  JHEP {\bf 1208}, 028 (2012)
  [arXiv:1112.2206 [hep-ph]];
M.~Adeel Ajaib, I.~Gogoladze, Q.~Shafi and C.~S.~Un,
  JHEP {\bf 1307}, 139 (2013)
  [arXiv:1303.6964 [hep-ph]].

\bibitem{Amhis:2012bh} 
  Y.~Amhis {\it et al.} [Heavy Flavor Averaging Group],
  arXiv:1207.1158 [hep-ex].
  
\bibitem{Misiak:2006zs} 
  M.~Misiak {\it et al.},
  Phys.\ Rev.\ Lett.\  {\bf 98}, 022002 (2007)
  doi:10.1103/PhysRevLett.98.022002
  [hep-ph/0609232].

\bibitem{Deshpande:1987nr} 
  N.~G.~Deshpande, P.~Lo, J.~Trampetic, G.~Eilam and P.~Singer,
  Phys.\ Rev.\ Lett.\  {\bf 59}, 183 (1987).
  doi:10.1103/PhysRevLett.59.183;
  S.~Bertolini, F.~Borzumati and A.~Masiero,
  Phys.\ Rev.\ Lett.\  {\bf 59}, 180 (1987).
  doi:10.1103/PhysRevLett.59.180;
  J.~L.~Hewett,
  Phys.\ Lett.\ B {\bf 193}, 327 (1987).
  doi:10.1016/0370-2693(87)91245-7.
  T.~G.~Rizzo,
  Phys.\ Rev.\ D {\bf 38}, 820 (1988)
  doi:10.1103/PhysRevD.38.820.

\bibitem{Idarraga:2008zz} 
  J.~P.~Idarraga, R.~Martinez, J.~A.~Rodriguez and N.~Poveda,
  Braz.\ J.\ Phys.\  {\bf 38}, 531 (2008)
  doi:10.1590/S0103-97332008000500001.




\bibitem{Khachatryan:2014wca} 
  V.~Khachatryan {\it et al.} [CMS Collaboration],
  JHEP {\bf 1410}, 160 (2014)
  doi:10.1007/JHEP10(2014)160
  [arXiv:1408.3316 [hep-ex]].

\bibitem{Heinemeyer:2013tqa}
  S.~Heinemeyer {\it et al.} [LHC Higgs Cross Section Working Group],
  doi:10.5170/CERN-2013-004
  arXiv:1307.1347 [hep-ph].

\bibitem{Djouadi:2005gj} 
  A.~Djouadi,
  Phys.\ Rept.\  {\bf 459}, 1 (2008)
  doi:10.1016/j.physrep.2007.10.005
  [hep-ph/0503173].

\bibitem{Baglio:2015wcg} 
  J.~Baglio, A.~Djouadi and J.~Quevillon,
  Rept.\ Prog.\ Phys.\  {\bf 79}, no. 11, 116201 (2016)
  doi:10.1088/0034-4885/79/11/116201
  [arXiv:1511.07853 [hep-ph]].

\bibitem{Bennett:2006fi} 
  G.~W.~Bennett {\it et al.} [Muon g-2 Collaboration],
  Phys.\ Rev.\ D {\bf 73}, 072003 (2006)
  [hep-ex/0602035];
  G.~W.~Bennett {\it et al.} [Muon (g-2) Collaboration],
  Phys.\ Rev.\ D {\bf 80}, 052008 (2009)
  [arXiv:0811.1207 [hep-ex]].

\bibitem{Davier:2010nc} 
  M.~Davier, A.~Hoecker, B.~Malaescu and Z.~Zhang,
  Eur.\ Phys.\ J.\ C {\bf 71}, 1515 (2011)
  Erratum: [Eur.\ Phys.\ J.\ C {\bf 72}, 1874 (2012)]
  [arXiv:1010.4180 [hep-ph]];
  K.~Hagiwara, R.~Liao, A.~D.~Martin, D.~Nomura and T.~Teubner,
  J.\ Phys.\ G {\bf 38}, 085003 (2011)
  [arXiv:1105.3149 [hep-ph]].

  
\end{thebibliography}
\end{document}